\documentclass[11pt]{article}
\pdfoutput=1

\usepackage{jheppub}
\bibstyle{JHEP}

\usepackage[english]{babel} 
\usepackage[utf8]{inputenc}  
\usepackage[T1]{fontenc}
\usepackage{lmodern}
\usepackage{amsmath}         
\usepackage{amsfonts}          
\usepackage{amssymb}
\usepackage{graphicx} 
\usepackage{bbold}
\usepackage[colorlinks=true,urlcolor=blue,linkcolor=blue,citecolor=blue,linktocpage=true]{hyperref}

\newcommand{\NN}{{\mathbb{N}}}

\newcommand{\Tr}{{\mathrm{Tr}}}
\newcommand{\cA}{{\mathcal{A}}}
\newcommand{\cF}{{\mathcal{F}}}
\newcommand{\cH}{{\mathcal{H}}}

\newcommand{\cO}{{\mathcal{O}}}
\newcommand{\cS}{{\mathcal{S}}}
\newcommand{\cZ}{{\mathcal{Z}}}

\newcommand{\ket}[1] {{| #1 \rangle}}

\usepackage{tikz}
\usetikzlibrary{calc}

\begin{document}

\title{Illuminating entanglement shadows of BTZ black holes by a generalized entanglement measure}

\author{Marius Gerbershagen}
\affiliation{Institut f{\"u}r Theoretische Physik und Astrophysik \\ and W\"urzburg-Dresden Cluster of Excellence ct.qmat, \\ Julius-Maximilians-Universit{\"a}t W{\"u}rzburg, Am Hubland, 97074 W\"urzburg, Germany}
\emailAdd{marius.gerbershagen@physik.uni-wuerzburg.de}

\abstract{
  We define a generalized entanglement measure in the context of the AdS/CFT correspondence.
  Compared to the ordinary entanglement entropy for a spatial subregion dual to the area of the Ryu-Takayanagi surface, we take into account both entanglement between spatial degrees of freedom as well as between different fields of the boundary theory.
  Moreover, we resolve the contribution to the entanglement entropy of strings with different winding numbers in the bulk geometry.
  We then calculate this generalized entanglement measure in a thermal state dual to the BTZ black hole in the setting of the D1/D5 system at and close to the orbifold point.
  We find that the entanglement entropy defined in this way is dual to the length of a geodesic with non-zero winding number.
  Such geodesics probe the entire bulk geometry, including the entanglement shadow up to the horizon in the one-sided black hole as well as the wormhole growth in the case of a two-sided black hole for an arbitrarily long time.
  Therefore, we propose that the entanglement structure of the boundary state is enough to reconstruct asymptotically AdS$_3$ geometries up to extremal surface barriers.
}

\maketitle

\section{Introduction}
\label{sec:introduction}
In the AdS/CFT correspondence, an increasingly important role has been assumed by the question of how the bulk geometry is encoded in the boundary field theory.
The Ryu-Takayanagi formula \cite{Ryu:2006bv} provides a connection between the entanglement entropy $S_A$ of a spatial subregion $A$ on the boundary and a codimension two minimal surface $\gamma_A$ homologous to $A$ in the bulk,
\begin{equation}
  S_A = \frac{\text{Area}(\gamma_A)}{4 G_N}.
  \label{eq:RT-formula}
\end{equation}
This is a clear indication that the entanglement structure of the field theory state is vital to understanding the encoding between bulk geometry and boundary data, subsumed under the slogan ``entanglement builds geometry'' \cite{Swingle:2009bg,VanRaamsdonk:2010pw,Bianchi:2012ev}.

However, it is also known that the entanglement entropy between spatial subsets of the field theory degrees of freedom as measured by the RT formula \eqref{eq:RT-formula} is not enough to reconstruct the full bulk geometry.
In particular, there are so-called entanglement shadows \cite{Hubeny:2013gta,Balasubramanian:2014sra,Freivogel:2014lja} -- certain subregions of the bulk geometry close to black hole horizons or naked singularities -- which are not penetrated by any RT surface.
Thus, knowledge of the RT surfaces does not translate into knowledge about the spacetime inside the entanglement shadow, precluding attempts of reconstructing this part of the bulk geometry from entanglement data.
A particular striking example of an entanglement shadow occurs in a two-sided black hole.
In this case, the growth of the area of a Cauchy slice through the wormhole asymptoting to a fixed time $t$ on the two boundaries is only captured by the RT formula for a short period in $t$ \cite{Hartman:2013qma}, which has lead to the proposal that features other than the entanglement entropy of the boundary state are needed to describe the part of the wormhole geometry behind the horizon \cite{Susskind:2014moa,Susskind:2014rva,Stanford:2014jda}.

In this publication, we investigate a generalization of the RT formula capturing entanglement for a subset of the degrees of freedom that includes a part of the base space where the boundary theory is defined as well as the target space spanned by the fields of the boundary theory.
In other words, we are considering entanglement between both the spatial degrees of freedom and between different fields of the boundary theory.
The entanglement between such degrees of freedom has been studied previously for conical defects in AdS$_3$, where it has been given the name ``entwinement'' \cite{Balasubramanian:2014sra,Balasubramanian:2016xho,Balasubramanian:2018ajb,Erdmenger:2019lzr}.
In the conical defect case, it was found that entwinement is dual to the length of a non-minimal geodesic winding a certain number of times around the naked singularity in the bulk \cite{Balasubramanian:2014sra,Balasubramanian:2016xho}, in contrast to the RT geodesics for the ordinary entanglement entropy between spatial subregions which have zero winding number.
The non-minimal geodesics probe the entire bulk geometry including the entanglement shadow around the conical defect.

\begin{figure}
  \centering
  \begin{tikzpicture}[scale=0.3]
    \begin{scope}[shift={(-6,0)}]
      \draw (0,0) circle(5);
      \fill[gray] (0,0) circle(2.5);
      \fill[black] (0,0) circle(1);
      \draw[red] ({{5*cos(60)}},{{5*sin(60)}}) to[out=220,in=0] (0,-1.75);
      \draw[red] (0,-1.75) arc(-90:-270:1.75) (1.75,0);
      \draw[red] (0,1.75) to[out=0,in=140] ({{5*cos(60)}},{{-5*sin(60)}});
    \end{scope}
    \begin{scope}[shift={(6,0)}]
      \draw (0,0) circle(5);
      \fill[gray] (0,0) circle(2.5);
      \fill[black] (0,0) circle(1);
      \draw[red] ({{5*cos(60)}},{{5*sin(60)}}) to[out=240,in=90] (3.25,0);
      \draw[red] (3.25,0) to[out=-90,in=110] ({{5*cos(60)}},{{-5*sin(60)}});
    \end{scope}
  \end{tikzpicture}
  \caption{Geodesics winding around the black hole penetrate the entanglement shadow (shown in gray) while non-winding Ryu-Takayanagi geodesics with minimal length stay outside of the entanglement shadow.}
  \label{fig:winding-geodesics}
\end{figure}
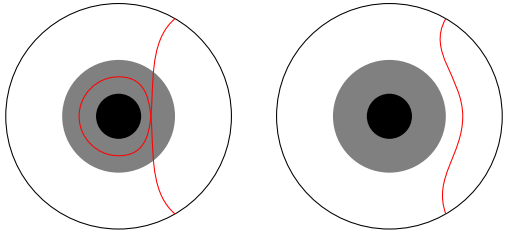

We study notions of entanglement for such non-spatially organized degrees of freedom in the case of thermal states dual to the BTZ black hole and thermal AdS$_3$.
Concretely, we work in the setting of the D1/D5 system (see \cite{David:2002wn} for a review).
We start our investigation at the orbifold point in the moduli space where the boundary CFT is given by the weakly coupled $S_N$ orbifold theory.
We use the following new ingredients to define our generalized entanglement measure compared to the ordinary entanglement entropy of spatial subregions as used in RT formula:
\begin{enumerate}
\item We resolve different twisted sectors.
  The $S_N$ orbifold theory includes sectors where the fields (collectively denoted by $X^1,...,X^N$) obey non-trivial boundary conditions $X^i(\phi + 2\pi) = X^{g(i)}(\phi)$ for some permutation $g \in S_N$.
  By projecting onto some subset of these sectors, we are able to resolve the contribution of states with particular boundary conditions which are dual to collections of string worldsheets with non-trivial winding numbers.
\item We consider entanglement for a subset of the degrees of freedom consisting of different fields localized in different subregions.
  For the ordinary entanglement entropy, the degrees of freedom considered consist of all fields $X^i$ localized in the same subregion $A$.
  Here, on the other hand, we consider different subregions $A^i$ for different fields $X^i$.
\end{enumerate}
By choosing the subregions $A^i$ and projections onto twisted sectors appropriately, we obtain an entanglement entropy result which is proportional to the length of a geodesic winding around the horizon of the BTZ black hole (see fig.~\ref{fig:winding-geodesics}).
Such winding geodesics penetrate the entanglement shadow, allowing for an explicit reconstruction of the entire bulk geometry up to the horizon from boundary data.
This generalizes to the two-sided case, where we find dual geodesics stretching between the two asymptotic boundaries which in the limit $N \to \infty$ can probe the entire wormhole geometry, including the region behind the horizon for an arbitrarily long time.
Finally, we comment on the string theory interpretation of our results and on deformations of the boundary CFT away from the orbifold point.

\section{$S_N$ orbifold at large $N$}
\label{sec:S_N-orbifold-general}
We begin by introducing the $S_N$ orbifold theory and reviewing the relevant features of this CFT which we will need in the following.
See for instance \cite{David:2002wn} and references therein for more details.

The $S_N$ theory is constructed by taking $N$ copies of a seed CFT with central charge $\tilde c$ \footnote{We use the convention that quantities with a tilde belong to the seed theory, while quantities without a tilde belong to the $S_N$ orbifold theory.} and identifying the copies under the $S_N$ permutation symmetry.
In the case of the D1/D5 system, the seed theory is given by a free supersymmetric CFT with target space $T^4$, however we will not need the specific form of the seed theory in the following and will thus leave it unspecified.
The total central charge is given by $c = N\tilde c$.
To obtain a modular invariant theory, it is necessary to introduce twisted sectors which implement non-trivial boundary conditions for the fields of the orbifold theory.
A twisted sector is labeled by a conjugacy class of $S_N$ corresponding to a collection of cycles along which the fields are fused together into a single strand by the boundary conditions.
% For instance, a single cycle of length 2 involving the fields $X^1$ and $X^2$ leads to the boundary conditions
% \begin{equation}
%   X^1(\phi + 2\pi) = X^2(\phi),~ X^2(\phi + 2\pi) = X^1(\phi).
% \end{equation}
Thus, a twisted sector is specified by the number $n_m$ of cycles of length $m$.
We denote such a twisted sector by $(1)^{n_1}(2)^{n_2}...(N)^{n_N}$.
The total length of the cycles is given by the number of copies of the seed theory,
\begin{equation}
  N = \sum_m m n_m.
\end{equation}

The thermal partition function of this theory is determined from the following recursion formula (see app.~\ref{sec:S_N-orbifold-partition-function}),
\begin{equation}
  Z_N(\tau) = \frac{1}{N} \sum_{k=1}^N \sum_{l=1}^{\lfloor N/k \rfloor} \sum_{j=0}^{k-1} \tilde Z\left(\frac{\tau l + j}{k}\right) Z_{N-kl}(\tau),
  \label{eq:S_N-orbifold-partition-function}
\end{equation}
where $\tilde Z(\tau)$ is the partition function of the seed theory.
In the following, we restrict to purely imaginary $\tau = \frac{i\beta}{2\pi}$.
At large $N$, the partition function \eqref{eq:S_N-orbifold-partition-function} is equal to the universal form valid for general two dimensional holographic CFTs \cite{Hartman:2014oaa}\footnote{In this publication, we take ``universal'' to mean dependent only on the central charge of the CFT and not on details of the spectrum or OPE coefficients.},
\begin{equation}
  \log Z_N(\tau) = \left\{
    \begin{aligned}
      \frac{\tilde c N}{12} \beta + \cO(N^0) &,& \beta > 2\pi\\
      \frac{\tilde c N}{12} \frac{4\pi^2}{\beta} + \cO(N^0) &,& \beta < 2\pi.\\
    \end{aligned}\right.
  \label{eq:universal-form-S_N-partition-function}
\end{equation}
This is derived in \cite{Hartman:2014oaa} by noting that $\tilde Z(\tau)$ can be bounded by
\begin{equation}
  \tilde Z(\tau) \leq p(\beta) \exp\left(\frac{\tilde c}{12} \beta\right)\exp\left(\frac{\tilde c}{6} \frac{4\pi^2}{\beta}\right),
  \label{eq:upper-bound-seed-partition-function}
\end{equation}
for any seed theory, where $p(\beta)$ is some polynomial in $\beta$.
Inserting this into \eqref{eq:S_N-orbifold-partition-function}, we see that at large $N$ the leading terms come from the $k=1,j=0$ terms if $\beta > 2\pi$ and the $l=1,j=0$ terms if $\beta < 2\pi$.
Other terms in $Z_N$ are exponentially suppressed with $e^{-N}$.

In the following, we will also need the decomposition of \eqref{eq:S_N-orbifold-partition-function} into twisted sectors.
The contribution of a single cycle of length $m$ is given by
\begin{equation}
  Z_{(m)} = \frac{1}{m} \sum_{j=0}^{m-1} \tilde Z\left(\frac{\tau + j}{m}\right).
  \label{eq:single-cycle-contribution}
\end{equation}
In general, a twisted sector contains multiple cycles which in total give a contribution to the partition function of
\begin{equation}
  Z_{(1)^{n_1}...(N)^{n_N}} = \prod_{m=1}^N \frac{1}{n_m} \sum_{k=1}^{n_m} Z_{(m)}(k\tau) Z_{(m)^{n_m-k}}(\tau),
  \label{eq:twisted-sector-contribution}
\end{equation}
where $Z_{(m)^{n_m-k}}(\tau)$ is the contribution of the $(m)^{n_m-k}$ twisted sector recursively determined from \eqref{eq:twisted-sector-contribution}.
The total partition function \eqref{eq:S_N-orbifold-partition-function} is given by summing over all sectors.
The dominant contribution to $Z_N$ comes from the untwisted $(1)^N$ sector for $\beta > 2\pi$, while for $\beta < 2\pi$ all sectors contribute.
This can be seen explicitly by comparing \eqref{eq:twisted-sector-contribution} with the dominant contributions to \eqref{eq:S_N-orbifold-partition-function} \footnote{Another argument can be given as follows.
  From \cite{Hartman:2014oaa}, it is known that the identity character $\chi_\mathbb{1}(\tau)$ belonging to the untwisted sector dominates the partition function for $\beta > 2\pi$, while for $\beta < 2\pi$, the modular transformed identity character $\chi_\mathbb{1}(-1/\tau)$ dominates.
  The modular transformation exchanges the time and space directions of the torus on which the CFT lives.
  The identity character includes a sum over spin structures with all possible $S_N$ boundary conditions along the time direction.
  Hence, the modular transformed identity character includes a sum over spin structures with untwisted boundary conditions in the time direction and all possible $S_N$ boundary conditions along the space direction.}.

\section{Entanglement in the $S_N$ orbifold theory}
\label{sec:entanglement-S_N-orbifold}
We now turn to the calculation of entanglement entropies in the $S_N$ orbifold theory.
Usually in the AdS/CFT context, entanglement entropy is defined by the von Neumann entropy of the reduced density matrix $\rho_A$ of a spatial subregion $A$,
\begin{equation}
  S_A = -\Tr(\rho_A \log\rho_A).
  \label{eq:ordinary-entanglement-entropy-definition}
\end{equation}
In contrast to \eqref{eq:ordinary-entanglement-entropy-definition}, we now project the partition function onto some subset $\cS$ of the twisted sectors and consider the entanglement between non-spatial degrees of freedom.

The projection onto the subset $\cS$ is defined as follows.
The thermal density matrix of the $S_N$ orbifold theory decomposes into a sum over twisted sector contributions,
\begin{equation}
  \rho = \bigoplus_{\{n_m\}} p_{\{n_m\}} \rho_{\{n_m\}},
  \label{eq:thermal-density-matrix-decomposition-twisted-sectors}
\end{equation}
where $\rho_{\{n_m\}}$ is a normalized density matrix ($\Tr\rho_{\{n_m\}} = 1$) for the $\{n_m\}$ twisted sector and $p_{\{n_m\}} \in [0,1]$.
We split up $\rho$ into a part belonging to a particular collection $\cS$ of twisted sectors and a part for the remainder $\bar\cS$,
\begin{equation}
  \rho = p_\cS \rho_\cS \oplus p_{\bar\cS} \rho_{\bar\cS},
\end{equation}
where again $\rho_{\cS}$ and $\rho_{\bar\cS}$ are normalized and $p_\cS$, $p_{\bar\cS}$ denote the corresponding probability factors.
We then choose a particular subset $\cA$ of the degrees of freedom in $\cS$ described below and trace out the complement $\bar\cA$ to obtain a reduced density matrix $\rho_{\cA,\cS} = \Tr_{\bar\cA}\rho_\cS$ \footnote{Formally, we first need to enlarge the Hilbert space to include non $S_N$ invariant states before taking the partial trace, because as typical for a gauge theory the Hilbert space of the $S_N$ orbifold does not factorize into a tensor product $\cH_\cA \otimes \cH_{\bar\cA}$ (see e.g.~\cite{Erdmenger:2019lzr}).}.
From this, we define the entanglement entropy
\begin{equation}
  S_{\cA,\cS} = - \Tr (\rho_{\cA,\cS} \log\rho_{\cA,\cS}).
  \label{eq:entanglement-entropy}
\end{equation}
In this section we only consider the projection onto a $\cS = S_n$ subset of the twisted sectors of the full $S_N$ group, while projections onto single twisted sectors are considered in app.~\ref{sec:S_N-orbifold-partition-function}.
See also app.~\ref{sec:probability-factors} for details on the probability factors $p_{\{n_m\}}$.

Explicitly, the $S_n$ subset we use is defined as follows.
Let $n \in \NN$ and consider the subset $\cS$ of twisted sectors containing only cycles whose length is a multiple of $\lfloor N/n \rfloor$.
Furthermore, we take $n$ to be proportional to $N$ in the limit $N \to \infty$, which we implicitly use in all of the following calculations.
In the case that $n$ is not a divisor of $N$, there will be $\cO(N^0)$ remaining cycles which can be chosen in an arbitrary way without influencing the leading order in $N$ of the entanglement measure we are computing.
Thus, in the following we will assume w.l.o.g.~that $n$ is a divisor of $N$.
The thermal partition function for the subset $\cS$ is then given by
\begin{equation}
  Z^\cS_n(\tau) = \frac{1}{n} \sum_{k=1}^n \sum_{l=1}^{\lfloor n/k \rfloor} \sum_{j=0}^{k-1} Z_{(N/n)}\left(\frac{\tau l + j}{k}\right) Z^\cS_{n-kl}(\tau)
  \label{eq:S_n-partition-function}
\end{equation}
where $Z_{(N/n)}$ is the contribution of a single cycle of length $N/n$ determined from \eqref{eq:single-cycle-contribution} \footnote{Note that even though $Z^\cS_n(\tau)$ takes on the form of a $S_n$ orbifold partition function, it is not modular invariant since the equivalent of the seed partition function, $Z_{(N/n)}$, in this case is not modular invariant.
  Since we are projecting onto a subset of the Hilbert space, this is expected and does not lead to any contradictions in the subsequent calculations.}.

What is left to do is to choose a subset $\cA$ of the degrees of freedom for which to calculate the entanglement entropy.
States of the $S_N$ orbifold theory are given by $S_N$ invariant combinations
\begin{equation}
  \frac{1}{N}\sum_{g \in S_N} g\ket\psi.
  \label{eq:S_N-invariant-state}
\end{equation}
Applying $g$ onto $\ket\psi$ leaves the number of cycles $n_m$ of the twisted sector to which $\ket\psi$ belongs invariant, but changes which fields $X^i$ are connected in a cycle.
Thus to specify the subset $\cA$, we choose first a fixed $S_N$ element $g_\psi$ giving the boundary conditions $X^i(\phi + 2\pi) = X^{g_\psi(i)}(\phi)$ for $\ket\psi$ and then second a spatial subregion $A^i$ for each $X^i$.
This is to be done separately for each twisted sector in $\cS$.
Proper $S_N$ invariant states in $\cA$ are then obtained from by applying \eqref{eq:S_N-invariant-state} for the states in the chosen Hilbert space subset.

\subsection{Single interval}
\begin{figure}
  \centering
  \begin{tikzpicture}[scale=0.4,style=thick]
    \begin{scope}
      \draw[red] (0,2) arc(90:-180:2);
      \draw (-2,0) arc(-180:-270:2);
      \draw (0,1.8) -- (0,2.2);
      \draw (0,-1.8) -- (0,-2.2);
      \draw (2,0) node[right] {\small{$X^1,$}};
      \draw (-2,0) node[left] {\small{$X^2$}};
    \end{scope}
    \begin{scope}[shift={(8,0)}]
      \draw[red] (0,2) arc(90:-180:2);
      \draw (-2,0) arc(-180:-270:2);
      \draw (0,1.8) -- (0,2.2);
      \draw (0,-1.8) -- (0,-2.2);
      \draw (2,0) node[right] {\small{$X^3$;}};
      \draw (-2,0) node[left] {\small{$X^4$}};
    \end{scope}
    \begin{scope}[shift={(18,0)},scale=1.2]
      \draw[red] (0,2) arc(90:-45:2);
      \draw ({{2*cos(-45)}},{{2*sin(-45)}}) arc(-45:-90:2);
      \draw[red] (0,-2) arc(-90:-225:2);
      \draw ({{2*cos(-225)}},{{2*sin(-225)}}) arc(-225:-270:2);
      \draw (0,1.8) -- (0,2.2);
      \draw (1.8,0) -- (2.2,0);
      \draw (0,-1.8) -- (0,-2.2);
      \draw (-1.8,0) -- (-2.2,0);
      \draw ({{2*cos(45)}},{{2*sin(45)}}) node[right,above] {\small{$X^1$}};
      \draw ({{2*cos(-45)}},{{2*sin(-45)}}) node[right,below] {\small{$X^2$}};
      \draw ({{2.2*cos(-135)}},{{2.2*sin(-135)}}) node[left,below] {\small{$X^3$}};
      \draw ({{2*cos(-225)}},{{2*sin(-225)}}) node[left,above] {\small{$X^4$}};
    \end{scope}
    \draw (10,-4) node {generalized entanglement entropy};
  \end{tikzpicture}\\[3mm]
  \begin{tikzpicture}[scale=0.4,style=thick]
    \begin{scope}[scale=0.6]
      \foreach \i in {1,2,3,4} {
        \draw[red] ({{5*(\i-1)}},2) arc(90:-90:2);
        \draw ({{5*(\i-1)}},-2) arc(-90:-270:2);
        \draw ({{5*(\i-1)}},-2) node[below] {\small{$X^\i$}};
        \draw ({{5*(\i-1)}},1.7) -- ({{5*(\i-1)}},2.3);
      }
    \end{scope}
    \begin{scope}[scale=0.8,shift={(2,-6)}]
      \draw[red] (0,2) arc(90:0:2);
      \draw (2,0) arc(0:-90:2);
      \draw[red] (0,-2) arc(-90:-180:2);
      \draw (-2,0) arc(-180:-270:2);
      \draw (0,1.8) -- (0,2.2);
      \draw (0,-1.8) -- (0,-2.2);
      \draw (2,0) node[right] {\small{$X^1,$}};
      \draw (-2,0) node[left] {\small{$X^2$}};
    \end{scope}
    \begin{scope}[scale=0.6,shift={(1,-8)}]
      \foreach \i in {3,4} {
        \draw[red] ({{5*(\i-1)}},2) arc(90:-90:2);
        \draw ({{5*(\i-1)}},-2) arc(-90:-270:2);
        \draw ({{5*(\i-1)}},-2) node[below] {\small{$X^\i$}};
        \draw ({{5*(\i-1)}},1.7) -- ({{5*(\i-1)}},2.3);
      }
    \end{scope}
    \begin{scope}[scale=0.8,shift={(21,0)}]
      \draw[red] (0,2) arc(90:0:2);
      \draw (2,0) arc(0:-90:2);
      \draw[red] (0,-2) arc(-90:-180:2);
      \draw (-2,0) arc(-180:-270:2);
      \draw (0,1.8) -- (0,2.2);
      \draw (0,-1.8) -- (0,-2.2);
      \draw (2,0) node[right] {\small{$X^1,$}};
      \draw (-2,0) node[left] {\small{$X^2$}};
    \end{scope}
    \begin{scope}[scale=0.8,shift={(30,0)}]
      \draw[red] (0,2) arc(90:0:2);
      \draw (2,0) arc(0:-90:2);
      \draw[red] (0,-2) arc(-90:-180:2);
      \draw (-2,0) arc(-180:-270:2);
      \draw (0,1.8) -- (0,2.2);
      \draw (0,-1.8) -- (0,-2.2);
      \draw (2,0) node[right] {\small{$X^3,$}};
      \draw (-2,0) node[left] {\small{$X^4$}};
    \end{scope}
    \begin{scope}[shift={(18,-5)}]
      \draw[red] (0,2) arc(90:30:2);
      \draw ({{2*cos(30)}},{{2*sin(30)}}) arc(30:-30:2);
      \draw[red] ({{2*cos(-30)}},{{2*sin(-30)}}) arc(-30:-90:2);
      \draw ({{2*cos(-90)}},{{2*sin(-90)}}) arc(-90:-150:2);
      \draw[red] ({{2*cos(-150)}},{{2*sin(-150)}}) arc(-150:-210:2);
      \draw ({{2*cos(-210)}},{{2*sin(-210)}}) arc(-210:-270:2);
      \draw (0,1.8) -- (0,2.2);
      \draw ({{1.8*cos(-30)}},{{1.8*sin(-30)}}) -- ({{2.2*cos(-30)}},{{2.2*sin(-30)}});
      \draw ({{1.8*cos(-150)}},{{1.8*sin(-150)}}) -- ({{2.2*cos(-150)}},{{2.2*sin(-150)}});
      \draw ({{2.3*cos(30)}},{{2.3*sin(30)}}) node[right,above] {\small{$X^1$}};
      \draw ({{2*cos(-90)}},{{2*sin(-90)}}) node[below] {\small{$X^2$}};
      \draw ({{2.2*cos(-210)}},{{2.2*sin(-210)}}) node[left,above] {\small{$X^3$}};
    \end{scope}
    \begin{scope}[scale=0.6,shift={(38,-8)}]
      \draw[red] (0,2) arc(90:-90:2);
      \draw (0,-2) arc(-90:-270:2);
      \draw (0,-2) node[below] {\small{$X^4$}};
      \draw (0,1.7) -- (0,2.3);
    \end{scope}
    \draw (26.5,-5) node {etc.};
    \draw (15,-9) node {ordinary entanglement entropy};
  \end{tikzpicture}
  \caption{Boundary conditions and degrees of freedom for the ordinary and generalized entanglement entropy in the example of $N=4$, $n=2$ and $\hat k=1$.
    A circle in the figure with $m$ field labels attached to it corresponds to a cycle of length $m$ in the group element defining the twisted boundary conditions.
    The red intervals mark the degrees of freedom for which the entanglement entropy is defined.
    For the ordinary entanglement entropy, we consider all twisted sectors and all fields on the same interval $[0,z_2]$.
    For the generalized entanglement entropy, we consider only cycles of length $m \in 2\NN$ and the intervals as given as $[0,2\pi]$ for $X^{1,3}$ and $[0,z_2]$ for $X^{2,4}$.
    The boundary conditions effectively connect the intervals for the $X^1,X^2$ as well as the $X^3,X^4$ fields into a single interval of size $2\pi \hat k + z_2$.
    Note that we draw only one realization of each twisted sector.
    Symmetrization over the $S_N$ group would yield further realizations by exchanging the field labels in the above picture.
  }
  \label{fig:field-theory-DoF}
\end{figure}
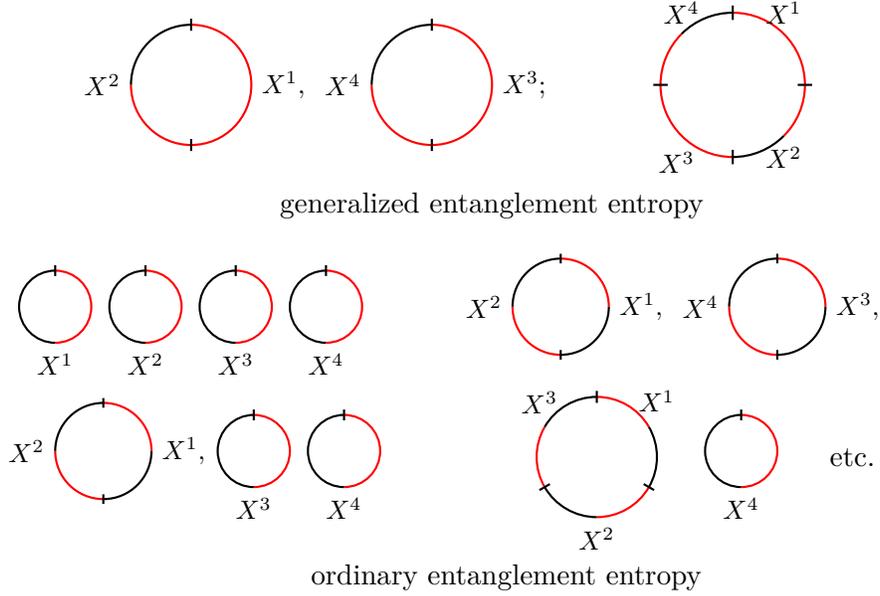

We first consider a subset consisting of $\hat k$ fields $X^i(\phi \in [0,2\pi]),...,X^{i+\hat k-1}(\phi \in [0,2\pi])$ in the full space together with a single field $X^{i+\hat k}(\phi \in [0,z_2])$ in some spatial subregion $[0,z_2]$ where $i \in (N/n) \NN$.
The $S_N$ element $g_\psi$ is given such that all of the fields in $\cA$ are continuously connected by the twisted boundary conditions in the sense that $X^{i+j}(\phi + 2\pi) = X^{i+(j+1)\text{mod}(N/n)}(\phi)$ for all $j \in \{0,...,N/n-1\}$.
We refer to this choice of degrees of freedom as the single interval case, since the boundary conditions effectively connect $N/n$ fields together into a single field, where we consider the degrees of freedom of a single interval touching $\hat k+1$ fields (see fig.~\ref{fig:field-theory-DoF}).

The entanglement entropy \eqref{eq:entanglement-entropy} is obtained via the replica trick \cite{Calabrese:2004eu}: we first calculate the Rényi entropy
\begin{equation}
  S^{(\alpha)}_{\cA,\cS} = \frac{1}{1-\alpha} \log\Tr\rho_{\cA,\cS}^\alpha
  \label{eq:Renyi-entropy}
\end{equation}
for integer $\alpha$ and then analytically continue to $\alpha \to 1$ to obtain \eqref{eq:entanglement-entropy}.
The Rényi entropy is in turn obtained from the partition function $Z^\cS_{\alpha,\text{replica}}(\cA)$ on a higher genus surface constructed by gluing together $\alpha$ copies of the system along the entangling interval determined by $\cA$,
\begin{equation}
  S_{\cA,\cS} = - \lim_{\alpha \to 1} \partial_\alpha \frac{Z^\cS_{\alpha,\text{replica}}(\cA)}{(Z^\cS_{1,\text{replica}}(\cA))^\alpha}.
  \label{eq:replica-trick}
\end{equation}
The replica partition function $Z^\cS_{\alpha,\text{replica}}(\cA)$ decomposes into conformal blocks,
\begin{equation}
  Z^\cS_{\alpha,\text{replica}}(\cA) = \sum_{p,q} a_{p,q} \cF_\alpha(h_p,h_q;\cA) \bar\cF_\alpha(h_p,h_q;\cA),
\end{equation}
where $a_{p,q}$ denotes the contribution of the OPE coefficients and multiplicities.
At large $c$, the conformal blocks exponentiate: $\cF(h_p,h_q;\cA) = e^{-c/6 f_\text{cl.}(h_p/c,h_q/c;\cA)}$.

For the ordinary entanglement entropy of a spatial subregion defined by \eqref{eq:ordinary-entanglement-entropy-definition}, the calculation proceeds as follows: the replica partition function $Z^{S_N}_{\alpha,\text{replica}}(A)$ is dominated by the identity zero-point conformal block on the higher genus replica surface at low temperature and its modular transformation $\tau \to -1/\tau$ at high temperature \cite{Gerbershagen:2021yma}.
By identity block we mean a conformal block whose internal operators all have vanishing conformal weight, $h_p=h_q=0$.
Contributions from other conformal blocks are suppressed by $e^{-c}$ factors because $f_\text{cl.}$ is an increasing function with increasing $h_p/c,h_q/c$ and $a_{p,q}$ grows subexponentially with $c$ \cite{Hartman:2013mia}\footnote{For exponentially growing $a_{p,q}$, the conformal block will still be dominant for smalls intervals, but the vacuum block dominance argument cannot exclude phase transitions at some finite value of the interval size \cite{Hartman:2013mia}.}.
This statement is known as vacuum block dominance and holds for CFTs with large central charge $c$, a sparse spectrum of light operators and at most exponentially growing OPE coefficients \cite{Hartman:2013mia}, including the $S_N$ orbifold studied here \cite{Hartman:2014oaa}.
The semiclassical conformal blocks are obtained from the solution of an auxiliary differential equation via a monodromy method \cite{Hartman:2013mia,Gerbershagen:2021yma}, yielding
\begin{equation}
  S_A = 
  \left\{
    \begin{aligned}
      \frac{c}{3} \log\left(\frac{\beta}{2\pi\epsilon}\sinh\left(\frac{2\pi^2(z_2 - z_1)}{\beta}\right)\right) &,& \beta < 2\pi\\
      \frac{c}{3} \log\left(\frac{1}{\epsilon}\sin\left(\pi(z_2-z_1)\right)\right) &,& \beta > 2\pi
    \end{aligned}
  \right.
  \label{eq:EE-result}
\end{equation}
for an entangling interval $A=[z_1,z_2]$ and UV cutoff $\epsilon$ \footnote{There is also a phase transition for high temperatures and large entangling intervals which we haven't shown in \eqref{eq:EE-result}.}.

The calculation of the entanglement entropy \eqref{eq:entanglement-entropy} considered here is then a straightforward generalization of the above calculation.
We refer to \cite{Hartman:2013mia,Gerbershagen:2021yma} for a more detailed explanation of the procedure and only explain the differences to the ordinary entanglement entropy of a spatial subregion in following.

\subsubsection*{Low temperatures}
The first important difference is that in the sum over states in the thermal partition function, the untwisted sector and thus the character for the identity operator is projected out for the subset $\cS$.
Hence, by the vacuum block dominance argument, the leading contribution to the thermal partition function at low temperature comes from the character of the operator with lowest conformal weight that is not projected out.
In our case this is the twist operator $\Sigma$ creating the ground state of the $(N/n)^n$ twisted sector with conformal weight
\begin{equation}
  h_\Sigma = \bar h_\Sigma = \frac{n\tilde c}{24} \left(\frac{N}{n} - \frac{n}{N}\right).
  \label{eq:lowest-conformal-weight-S_n-subset}
\end{equation}
We confirm this by a direct calculation of the thermal partition function of the $S_n$ subset in app.~\ref{sec:S_N-orbifold-partition-function}.

Therefore, at low temperature, the leading contribution to the replica partition function for the $S_n$ subset comes from the zero-point block where the internal operator originating from the sum over states in the thermal partition function is given by $\Sigma$ instead of the identity.
The projection onto the $S_n$ subset of twisted sectors does not spoil the argument that this conformal block is dominant up to $e^{-c}$ corrections since projecting out a part of the spectrum can only decrease the multiplicities and does not change the OPE coefficients.
Note that all descendants of a primary operator $\Sigma$ are in the same twisted sector as $\Sigma$, thus the projection does not change the conformal blocks themselves.
The modification of the entangling intervals also necessitates a change in the monodromy conditions used to derive the conformal block in \cite{Gerbershagen:2021yma}: instead of trivial monodromy around an entangling interval $A = [0,z_2]$ we impose trivial monodromy around a path encircling $\hat k$ times the spatial circle combined with the interval $[0,z_2]$.
Due to the projection onto the subset $\cS$ of twisted sectors considered, this choice of monodromy conditions is well defined since all of the fields touched by this path are sewn together into a continuous cycle by the twisted boundary conditions.

To compute the entanglement entropy, we then need to derive the zero-point conformal block on the replica surface for internal operators $\mathbb{1}$ and $\Sigma$.
This may be achieved in a perturbation expansion in $\alpha-1$ using the monodromy method derived in \cite{Gerbershagen:2021yma}.
The zeroth order in $\alpha-1$ gives the $h_\Sigma$ character as expected.
The first order in $\alpha-1$ yields the semiclassical zero-point block on the replica surface, analytically continued in $\alpha$,
\begin{equation}
  f_1 = \frac{n}{N} \log(\sin(\pi(z_2 + \hat k)n/N)),
  \label{eq:low-temperature-first-order-block}
\end{equation}
related to $Z^\cS_{\alpha,\text{replica}}$ by
\begin{equation}
  Z^\cS_{\alpha,\text{replica}} \propto \chi_\Sigma(\tau)\bar\chi_\Sigma(\bar\tau) e^{-c/6 (\alpha-1) (f_1 + \bar f_1)}.
\end{equation}
The proportionality constant includes OPE coefficients and multiplicity factors which drop out in the end.
Thus we find from \eqref{eq:replica-trick}
\begin{equation}
  S_{\cA,\cS} = \frac{n\tilde c}{3} \log\biggl[\frac{N}{n\epsilon} \sin\bigl(\frac{\pi(z_2 + \hat k)n}{N}\bigr) \biggr] ~~ (\beta > 2\pi\frac{N}{n}).
  \label{eq:entwinement-low-temperature}
\end{equation}
The crossover point $\beta = 2\pi N/n$ between the low and high temperature limits can be derived from the thermal partition function (see \eqref{eq:universal-S_n-partition-function}).
Eq.~\eqref{eq:entwinement-low-temperature} is proportional to the length of a geodesic in thermal AdS$_3$ with opening angle $2\pi(z_2 + \hat k)n/N$.
Note that the proportionality constant between the geodesic length and the entanglement entropy is smaller by a factor of $n/N$ compared to the RT formula \eqref{eq:RT-formula} due to there being only $n$ branch cuts in the replica surface in total, compared to $N$ branch cuts in the ordinary entanglement entropy of spatial subregions.

\subsubsection*{High temperatures}
In the high temperature case, the leading contribution to the replica partition function comes from the identity block like in the case of the ordinary entanglement entropy of a spatial subregion.
This can be seem from the fact that the modular transformed identity character $\chi_\mathbb{1}(-1/\tau)$ includes contributions from all twisted sectors, which implies that the high temperature result for the thermal partition function of the $S_n$ subset is equal to the one for the full $S_N$ theory -- up to a proportionality constant that drops out for the entanglement entropy in normalizing the reduced density matrix to one.
Hence, the entanglement entropy for the $S_n$ subset at high temperature is also given by a modular transformation of the identity block like the entanglement entropy for the full $S_N$ orbifold.
Differences between the two quantities come a different subregion $\cA$ and thus a different monodromy condition: as at low temperatures, we impose trivial monodromy around a path encircling $\hat k$ times the spatial circle combined with the interval $[0,z_2]$.

We obtain at high temperatures and for small intervals
\begin{equation}
  S_{\cA,\cS} = \frac{n\tilde c}{3} \log\biggl[\frac{\beta}{2\pi\epsilon} \sinh\bigl(\frac{2\pi^2(z_2+\hat k)}{\beta}\bigr)\biggr]  ~~ (\beta < 2\pi\frac{N}{n}).
  \label{eq:entwinement-high-temperature}
\end{equation}
This is proportional to the length of a geodesic in the BTZ geometry with opening angle $2\pi z_2$ and winding number $\hat k$, again with a proportionality factor $n/N$ times smaller than that of the RT formula \eqref{eq:RT-formula}.
Note that for $2\pi N/n > \beta > 2\pi$, the dual geometry is in fact thermal AdS$_3$ while we still obtain an entanglement entropy dual to the length of a geodesic in the BTZ black hole.

For large intervals, there is phase transition as in the ordinary entanglement entropy case in \cite{Gerbershagen:2021yma},
\begin{equation}
  S_{\cA,\cS} = \frac{N\tilde c}{3} \frac{2\pi^2}{\beta} + \frac{n\tilde c}{3} \log\biggl[\frac{\beta}{2\pi\epsilon} \sinh\bigl(\frac{2\pi^2(N/n - z_2-\hat k)}{\beta}\bigr)\biggr].
  \label{eq:entwinement-large-interval-high-temperature}
\end{equation}
Eq.~\eqref{eq:entwinement-large-interval-high-temperature} is dual to the thermal entropy of the BTZ black hole plus the length of a geodesic in the black hole geometry with opening angle $2\pi(1-z_2)$ and winding number $N/n-\hat k-1$.
Under the assumption that vacuum block dominance holds for all values of $z_2$ and $\hat k$, the entanglement entropy is given by the minimum of \eqref{eq:entwinement-high-temperature} and \eqref{eq:entwinement-large-interval-high-temperature} with a sharp crossover point.

\subsection{Multiple intervals and two-sided black holes}
Let us now consider a subset $\cA$ of the degrees of freedom consisting of multiple disconnected components along a continuously connected cycle of $N/n$ fields.
We specify this subset by a collection of an even number of coordinates $z_a \in [0,N/n]$ with $z_a < z_{a+1}$.
If the coordinate $z = \phi + j$ belonging to the field $X^{i+j}$ for some $i \in (N/n)\NN$ is contained in one of the intervals $[z_{2a},z_{2a-1}]$, then the degrees of freedom of the field $X^{i+j}$ at the coordinate $\phi$ belong to the subset $\cA$ in consideration.
The arguments given above then generalize to the multiple interval case in the same way as for the ordinary entanglement entropy in \cite{Gerbershagen:2021yma}, yielding
\begin{equation}
  S_{\cA,\cS} = \frac{n\tilde c}{3} \sum_{(i,j)} \log\biggl[\frac{N}{n\epsilon}\sin\bigl(\pi(z_i-z_j)\frac{N}{n}\bigr)\biggr]
  \label{eq:entwinement-multiple-intervals-low-temperature}
\end{equation}
for low temperature ($\beta < 2\pi N/n$) and
\begin{equation}
  S_{\cA,\cS} = \frac{n\tilde c}{3} \sum_{(i,j)} \log\biggl[\frac{\beta}{2\pi\epsilon}\sinh\bigl(\frac{2\pi^2(z_i-z_j)}{\beta}\bigr)\biggr]
  \label{eq:entwinement-multiple-intervals-high-temperature}
\end{equation}
for high temperature ($\beta > 2\pi N/n$) and small intervals.
Moreover, there is a phase transition for high temperatures and large intervals analogous to \eqref{eq:entwinement-large-interval-high-temperature}.
Which combination of pairs $(i,j)$ to take depends on the interval sizes.
As for the ordinary entanglement entropy, the combination that gives the lowest entanglement entropy dominates the replica partition function $Z^\cS_{\alpha,\text{replica}}$ if the vacuum block dominance property holds.

Of particular interest is the case of two intervals on opposite sides of a two-sided black hole.
This is obtained by placing one of the two intervals at an offset $+\tau/2$ on the torus of the boundary theory.
For two intervals of equal size $L < N/n$, we obtain the entanglement entropy
\begin{equation}
  S_{\cA,\cS} = \left\{
    \begin{aligned}
      \frac{2n\tilde c}{3} \log\biggl[\frac{\beta}{2\pi\epsilon}\cosh\bigl(\frac{4\pi^2t}{\beta}\bigr)\biggr] &,& t < t_c\\
      \frac{2n\tilde c}{3} \log\biggl[\frac{\beta}{2\pi\epsilon}\sinh\bigr(\frac{2\pi^2L}{\beta}\bigr)\biggr] &,& t > t_c\\
    \end{aligned}
  \right.
  \label{eq:entwinement-two-sided-black-hole}
\end{equation}
where $t_c = \frac{\beta}{4\pi^2}\text{arcosh}\sinh(2\pi^2 L/\beta) \approx L/2$.
For early times $t < t_c$, eq.~\eqref{eq:entwinement-two-sided-black-hole} is proportional to the length of two geodesics stretching from the endpoints of the entangling interval through the wormhole to the other side, while for late times $t > t_c$, the dual picture is given by two geodesics that do not enter the wormhole and wind $\lfloor L \rfloor$ times around the horizon.
Thus, the entanglement entropy for the degrees of freedom considered here can probe the wormhole growth of the BTZ black hole for a time $t_c \approx L/2$ that can be much larger than the one for the ordinary entanglement entropy of a spatial subregion which is restricted to $L < 1$.
For $N = \infty$, $t_c$ can be arbitrarily large\footnote{For finite but large $N$, eq.~\eqref{eq:entwinement-two-sided-black-hole} is valid in a series expansion in $N$ to the leading order if $n=\cO(N)$ and $L = \cO(N^0)$.
  To probe the wormhole growth for longer times in this case, it would be necessary to consider projections onto a smaller subset of the twisted sectors, for which the calculation methods used here no longer work.
  In \cite{Susskind:2014moa,Susskind:2014rva,Stanford:2014jda}, it was proposed that for finite $N$ the growth of the wormhole should continue for a time exponential in $N$.
  Naively extrapolating our calculation for the $S_n$ subset for large $n$ to the case $n=1$ corresponding to projecting onto the maximally twisted sector, it seems that the entanglement entropy will probe the wormhole growth only for a time that is linear in $N$.}.

\section{String theory interpretation}
\label{sec:string-theory-interpretation}
In this section, we explain the interpretation of the projection onto twisted sectors in the dual string theory picture.
At the orbifold point, the string theory is in the tensionless limit \cite{Giribet:2018ada,Gaberdiel:2018rqv,Eberhardt:2018ouy,Eberhardt:2019ywk,Dei:2020zui}.
In this limit, the string theory partition function on a thermal AdS$_3$ resp.~BTZ black hole background was derived in \cite{Eberhardt:2020bgq} and shown to agree with the orbifold partition function \eqref{eq:S_N-orbifold-partition-function}.
The leading contributions in the large $N$ limit come from spherical and toroidal string worldsheets \cite{Eberhardt:2020bgq}.
Moreover, the moduli of the torus worldsheets localize on holomorphic covering spaces of the boundary torus, meaning that the only contribution of torus worldsheets comes from strings that wind an integer number of times around the time or space circle of the boundary torus \cite{Eberhardt:2020bgq}.

From the equality of partition functions, it can be seen that each cycle of length $m$ in a twisted sector of the boundary theory corresponds to a string with winding number $m$ around the space circle of the boundary torus.
Therefore, the restriction to certain twisted sectors in the boundary theory amounts to considering only string worldsheets with particular winding numbers around the spatial circle.
For example, the untwisted sector corresponds to $N$ strings winding once around the spatial circle.
The $S_n$ subset considered in sec.~\ref{sec:entanglement-S_N-orbifold} corresponds to projecting onto the gravity degrees of freedom consisting of strings with winding number $k N/n$ for $k \in \NN$.

This makes it clear why the projection onto twisted sectors in the boundary theory leads to geodesics with non-zero winding number in the bulk: we are only considering strings with non-zero winding numbers and large enough subsets of the degrees of freedom on these strings such that we probe the winding around the non-contractible cycle in the bulk geometry.
We note, however, that of course the classical geometric description of the bulk spacetime for which the notion of a geodesic makes sense is not a good description in the tensionless limit and thus this interpretation is somewhat limited in its applicability.

\section{Deformations of the $S_N$ orbifold}
\label{sec:deformations}
At the orbifold point, the boundary theory is weakly coupled dual to a strongly coupled gravity theory.
In this section, we investigate which properties of the entanglement entropy expressions derived in sec.~\ref{sec:entanglement-S_N-orbifold} survive the deformation away from the orbifold point\footnote{For previous work on deformations of the $S_N$ orbifold, see \cite{Avery:2010er,Avery:2010hs,Pakman:2009mi,Asplund:2011cq,Burrington:2012yq,Carson:2014ena,Carson:2014yxa,Carson:2014xwa,Carson:2015ohj,Gaberdiel:2015uca,Carson:2016cjj,Carson:2016uwf,Burrington:2017jhh,Hampton:2018ygz,Guo:2019pzk,Guo:2019ady,Guo:2020gxm,Lima:2021wrz}.}.

The $S_N$ orbifold theory possesses 20 exactly marginal operators which can be used to deform the boundary theory to another CFT \cite{David:2002wn}.
16 of those operators are in the untwisted sector.
These are simple to handle since they leave the orbifold structure invariant: they just deform the CFT to a $S_N$ orbifold of a different seed theory.
But since the results of sec.~\ref{sec:entanglement-S_N-orbifold} are independent of the seed theory, these deformations do not change the entanglement entropy.

The 4 exactly marginal two-cycle twist operators are more interesting because they deform the CFT away from the weak coupling point \cite{David:2002wn,Gaberdiel:2015uca}.
In the following, we derive how this deformation affects the entanglement entropy to the first non-trivial order in conformal perturbation theory and at leading order in large $N$.
The derivation is based on the assumption that the deformed theory possesses the same vacuum block dominance properties (i.e.~sparse spectrum of low dimension operators and at most exponentially growing OPE coefficients) as the $S_N$ orbifold.
The fact that the RT formula at strong coupling gives the same entanglement entropy result \eqref{eq:EE-result} as we obtain at the weakly coupled orbifold point indicates that this assumption is likely justified, although we have not proven it from first principles.
Under this assumption, the entanglement entropy is obtained from the conformal block whose internal operators have the lowest dimension compatible with the projection onto the subset $\cS$ of twisted sectors, since we project out the same operators in the deformed theory as in the $S_N$ orbifold.

Therefore, it remains to determine the anomalous dimensions of the primaries of the $S_N$ orbifold theory.
The conformal weight of the identity operator in the deformed theory is given by $h=\bar h=0$ as in the $S_N$ orbifold.
Moreover, in app.~\ref{sec:anomalous-dimensions}, we argue that the anomalous dimension of operators in twisted sectors that consist of $n = \cO(N)$ cycles of length $\lfloor N/n \rfloor$ together with an arbitrary collection of cycles for the remaining $N - n$ fields vanishes up to $\cO(N^0)$ corrections to second order in conformal perturbation theory.
Under the assumption of vacuum block dominance, this implies that the partition functions considered in sec.~\ref{sec:entanglement-S_N-orbifold} are invariant under deformations up to $\cO(N^0)$ corrections.
This holds for both the thermal partition function $Z^\cS_n$ function for the $S_n$ subset as well as for the replica partition function $Z^\cS_{\alpha,\text{replica}}$ since these partition functions are dominated by characters or conformal blocks of the aforementioned operators which only receive small corrections to the conformal weight in the large $N$ limit\footnote{For the high temperature phase, this argument assumes that the projection onto the $S_n$ subset in the deformed theory does not project out exactly those twisted sectors that are responsible for the leading order contribution to the thermal partition function at high temperatures. It seems unlikely for this to happen, because projections with different $n$ project out different subsets of the twisted sectors and thus the leading order for the thermal partition function would have to come from the contributions of a very small subset of the twisted sectors that includes neither the vacuum (which we know not to be dominating for high temperature) nor any of the sectors that we project onto for some $n$. Furthermore, the results of sec.~\ref{sec:S_N-orbifold-partition-function} show that this does not occur for the $S_N$ orbifold theory.}.
Note that while the insertion of two-cycle twist operators into the replica partition function changes the boundary conditions for some of the fields of the $S_N$ orbifold and thus the monodromy conditions for the conformal blocks may differ from those of sec.~\ref{sec:entanglement-S_N-orbifold}, as long as we insert only a $\cO(N^0)$ number of such operators, this will not change the leading order in $N$ of the entanglement entropy result.

Thus we conclude that the entanglement entropy is invariant up to corrections of order $\cO(N^0)$ for deformations to second order in conformal perturbation theory.
It is likely that similar arguments can be used to show invariance also to higher orders, although the growing complexity of the involved conformal blocks and OPE coefficients quickly makes the calculation intractable.

\section{Discussion}
\label{sec:discussion}

The generalization of the RT formula to geodesics with non-zero winding number found in this publication considerably strengthens the ``entanglement builds geometry'' proposal in the AdS/CFT correspondence.
Due to the fact that the BTZ black hole is given as a quotient of pure AdS$_3$, the winding geodesics whose field theory dual we have found cover the entire black hole geometry up to the horizon in the one-sided case.
Moreover, in the two-sided case the non-minimal geodesics also include the wormhole geometry behind the horizon.
The correspondence between entanglement entropy and geodesic lengths then allows explicitly reconstructing the full bulk geometry from boundary data\footnote{For instance from integral geometry techniques which naturally incorporate winding geodesics \cite{Czech:2014wka,Czech:2014ppa,Czech:2015qta,Asplund:2016koz,Zhang:2016evx,Cresswell:2017mbk,Abt:2018ywl} or using the bit threads formulation \cite{Freedman:2016zud,Headrick:2017ucz,Cui:2018dyq,Agon:2020mvu}.}.
It is likely that the correspondence will also extend to other asymptotically AdS$_3$ spaces, since these geometries are all obtained as quotients of pure AdS$_3$ that naturally include winding geodesics.
This includes in particular conical defects, for which the correspondence between the entanglement entropy of non-spatial degrees of freedom and the length of winding geodesics has already been worked out under the catchphrase of ``entwinement'' \cite{Balasubramanian:2014sra,Balasubramanian:2016xho,Balasubramanian:2018ajb,Erdmenger:2019lzr}.

Hence, we conjecture that entanglement -- at least if one includes entanglement between non-spatial degrees of freedom and allows for resolving contributions of different twisted sectors -- is in fact enough to reconstruct the entire bulk geometry of asymptotically AdS$_3$ spaces up to extremal surface barriers.
Extremal surface barriers are codimension one surfaces which are not penetrated by any extremal surface of arbitrary codimension \cite{Engelhardt:2013tra}.
Examples of such surfaces include the black hole horizon for the one-sided BTZ black hole, which as we have seen is the deepest point in the bulk that the winding geodesics can reach.
This is not unexpected from since we are working in a thermal average on the boundary instead of a specific black hole microstate.

\medskip
Note that although our results were obtained at the orbifold point in the D1/D5 system, we expect them to hold more generally.
In particular, it is likely that the entanglement entropy results from sec.~\ref{sec:entanglement-S_N-orbifold} are invariant -- up to subleading corrections in the large central charge limit -- under deformations away from the weak coupling limit of the boundary theory, equivalent to turning on the string tension in the gravity theory.
For the ordinary entanglement entropy of a spatial subregion, which is a special case of our more general setup, this is certainly true given that the RT formula gives the same results in the strong coupling limit as what is obtained for the $S_N$ orbifold theory at weak coupling.
We explicitly confirmed the invariance of our generalized entanglement measure under deformations to the first non-trivial order in conformal perturbation theory in sec.~\ref{sec:deformations}.

Furthermore, we expect the interpretation of the restriction to string worldsheets with particular winding numbers as a resolution of particular (twisted) subsectors of the dual CFT to generalize to other top-down constructions of AdS$_3$/CFT$_2$ models.
This would allow one to define the entanglement measure proposed here for any AdS$_3$/CFT$_2$ model in which there exists a map between worldsheets with particular winding numbers and Hilbert space subsectors of the CFT.

\medskip
The generalized entanglement entropy defined in this work is related to other entanglement measures as follows.

First, it can be seen as a generalization of entwinement \cite{Balasubramanian:2014sra,Balasubramanian:2016xho,Balasubramanian:2018ajb,Erdmenger:2019lzr}.
While we discussed only thermal states, the procedure for calculating the generalized entanglement measure defined in this publication -- project onto a subset of the twisted sectors and then consider the entanglement entropy of non-spatially organized degrees of freedom -- works for any state in an orbifold theory.
For the special case of a pure state in a twisted sector containing only cycles of a fixed length, the resulting generalized entanglement measure is equal to entwinement\footnote{Of course, in this case the projection onto the subset of twisted sectors is trivial since the state is already in a single twisted sector.}.
In addition, taking the limit of the mass of the BTZ black hole going to zero, we find agreement with the results of \cite{Balasubramanian:2016xho}, where the generalized entanglement entropy for the massless BTZ black hole was calculated as a limit of the entwinement results for conical defects.

Second, let us also mention a similarity between our entanglement entropy definition and recent work on symmetry resolved entanglement \cite{Goldstein:2017bua,Xavier:2018kqb,Bonsignori:2019naz,Barghathi:2019oxr,Feldman:2019upn,Fraenkel:2019ykl,Tan:2019axb,Murciano:2020lqq,CapizziRuggieroCalabrese,Turkeshi:2020yxd,Murciano:2020vgh,Horvath:2020vzs,Azses:2020wfx,Zhao:2020qmn,Horvath:2021fks}.
Symmetry resolved entanglement is defined for a system with a global symmetry, for instance particle number conservation.
This symmetry induces a decomposition of reduced density matrices $\rho_A = \bigoplus_Q p_Q \rho_{A,Q}$ in subsectors of fixed subregion charge $Q$, i.e.~in the example of a conserved particle number the reduced density matrices decomposes into a sum over configurations with fixed number of particles in the subsystem $A$.
The symmetry resolved entanglement entropy $S_{A,Q}$ is then defined as the von Neumann entropy of the normalized density matrix $\rho_{A,Q}$.
The setup considered here is very similar, with the main difference being that in the symmetry resolved case, the decomposition into fixed charge subsectors happens only for the reduced density matrix while the total density matrix is assumed to have fixed total charge.
In contrast, in our case the density matrix \eqref{eq:thermal-density-matrix-decomposition-twisted-sectors} of the full system already decomposes into multiple sectors, which in turn leads to a similar decomposition of the reduced density matrix for the subsystem.
In our setup the analogue of the subregion charge $Q$ is the given configuration of strings with particular winding numbers corresponding to a conjugacy class of $S_N$.

\medskip
We close with mentioning possible future directions.
First of all, it would be interesting to investigate the correspondence between the entanglement of non-spatial degrees of freedom and geometric quantities in the bulk in higher dimensions.
While the situation in three dimensions is certainly special due to the BTZ black hole being a quotient of pure AdS$_3$, projections onto sectors of the boundary theory describing strings with certain winding numbers around black hole horizons or naked singularities might also be possible in higher dimensions.
A further point warranting further study is the possibility of finding a proof of the proposed relation between geodesic lengths and entanglement entropy in the spirit of \cite{Faulkner:2013yia,Lewkowycz:2013nqa}.
Moreover, it would be interesting to investigate whether analogues of entanglement wedges exist for the non-minimal winding geodesics in the bulk. 

\medskip\noindent
\textbf{Acknowledgments}\\
I would like to thank Johanna Erdmenger, Christian Northe, René Meyer and Suting Zhao for discussions.
I acknowledge financial support by the Deutsche Forschungsgemeinschaft (DFG, German Research Foundation) under Germany's Excellence Strategy through Würzburg‐Dresden Cluster of Excellence on Complexity and Topology in Quantum Matter ‐ ct.qmat (EXC 2147, project‐id 390858490).

\appendix

\section{Partition function of the $S_N$ orbifold}
\label{sec:S_N-orbifold-partition-function}
This appendix contains further results on the partition function of the $S_N$ orbifold theory.

\subsection{Recursion formula}
We first derive the recursion formula \eqref{eq:S_N-orbifold-partition-function} for the $S_N$ orbifold partition function $Z_N$ from its generating function $\cZ[p]$ \cite{Dijkgraaf:1996xw,Dijkgraaf:1998zd,Bantay:2000eq,Keller:2011xi},
\begin{equation}
  \cZ[p] = \sum_{N \geq 0} p^N Z_N(\tau) = \prod_{n > 0} \prod_{m,\bar m} (1-p^nq^{m/n}\bar q^{\bar m/n})^{-d(m,\bar m) \delta^{(n)}_{m-\bar m}}.
\end{equation}
Here,
\begin{equation}
  \delta^{(n)}_{m-\bar m} = \left\{
    \begin{aligned}
      1 &, ~~m-\bar m~\text{divides}~n \\
      0 &, ~~\text{otherwise.}
    \end{aligned}
  \right.
\end{equation}
The partition function is obtained by differentiating
\begin{equation}
  Z_N(\tau) = \left.\frac{1}{N!}\frac{\partial^N}{\partial p^N} \cZ[p] \right|_{p=0}.
\end{equation}
The first derivative w.r.t.~$p$ is given by
\begin{equation}
  \frac{\partial}{\partial p}\cZ = \sum_n\sum_{m,\bar m} \frac{n d(m,\bar m) \delta^{(n)}_{m-\bar m}}{p(p^{-n}q^{-m/n}\bar q^{-\bar m/n}-1)} \cZ.
\end{equation}
Using that for any $\alpha$,
\begin{equation}
  \frac{\partial^{l-1}}{\partial p^{l-1}} \left.\frac{1}{p(p^{-n}\alpha-1)}\right|_{p=0} = (l-1)! \sum_{k | l} \frac{\delta_{n,k}}{\alpha^{l/k}},
\end{equation}
we find the following recursion formula for the partition function,
\begin{equation}
  \begin{aligned}
    Z_N(\tau) &= \frac{1}{N} \sum_{l=1}^N \sum_{k | l} \sum_{j=0}^{k-1}\tilde Z\left(\tau\frac{l}{k^2}+\frac{j}{k}\right) Z_{N-l}(\tau).
  \end{aligned}
\end{equation}
The second sum in this formula is performed over all divisors $k$ of $l$.
Finally, we perform a resummation in $k$ such that all terms with the same $k$ are grouped together, yielding \eqref{eq:S_N-orbifold-partition-function}.

\subsection{$S_n$ partition function at large $n$}
We now derive the thermal partition function for the $S_n$ subset $\cS$ in the large $N$ limit.
Here, we give an argument from the decomposition of the partition function into characters before deriving the same results using the orbifold structure in the next subsection.

In general, the partition function of a large $c$ holographic CFT is obtained from the identity character $\chi_\mathbb{1}(\tau)$ at low temperature and its modular transformation $\chi_\mathbb{1}(-1/\tau)$ at high temperature \cite{Hartman:2014oaa,Gerbershagen:2021yma}.
That is, only descendants of the identity operator contribute to the sum over states in the partition function at low temperature.
As explained in sec.~\ref{sec:entanglement-S_N-orbifold}, the identity operator is projected out of the partition function for the subset $\cS$ and thus the leading contribution comes from the character of the operator $\Sigma$ with weight \eqref{eq:lowest-conformal-weight-S_n-subset}.
Thus, at low temperatures the partition function is dominated by the character
\begin{equation}
  \chi_\Sigma(\tau) \propto q^{h_\Sigma - c/24} = q^{-\frac{\tilde c}{24} \frac{n^2}{N}},
  \label{eq:dominating-character-S_n-subset}
\end{equation}
where $q=e^{2\pi i \tau} = e^{-\beta}$ and the proportionality constant includes only $\cO(N^0)$ factors.
At high temperatures, the partition function is dominated by the modular transformed vacuum character similar to the partition function for the full $S_N$ group, such that we obtain
\begin{equation}
  Z^\cS_n(\tau) \propto \left\{
    \begin{aligned}
      \exp\left(\frac{\tilde c}{12} \beta \frac{n^2}{N}\right) &,& \beta > 2\pi \frac{N}{n}\\
      \exp\left(\frac{\tilde c}{12} \frac{4\pi^2}{\beta} N\right) &,& \beta < 2\pi \frac{N}{n},\\
    \end{aligned}
  \right.
  \label{eq:universal-S_n-partition-function}
\end{equation}
where the proportionality constant includes a $n,N$ dependent prefactor for high temperatures.
This prefactor can not be derived by the above argument, however it may easily be obtained by inserting the results of the following subsection into \eqref{eq:S_n-partition-function}.

\subsection{Contribution of single twisted sectors at large $N$}
At large $N$, the partition function of the $S_N$ orbifold takes on the universal form \eqref{eq:universal-form-S_N-partition-function}.
This extends to the contribution of certain twisted sectors to the partition function.
In particular, we show that the contribution of a large number $n_m = \cO(N)$ of short cycles with length $m = \cO(N^0)$ is universal.
Similarly, the contribution of a small number of long cycles with length $m = \cO(N)$ is also universal in the large $N$ limit.

\subsubsection*{Short cycles}
Let us start with the case of the untwisted sector which includes $N$ cycles of length 1.
The corresponding contribution to the partition function is given by
\begin{equation}
  Z_{(1)^N}(\tau) = \frac{1}{N} \sum_{k=1}^N \tilde Z(k\tau) Z_{(1)^{N-k}}(\tau).
  \label{eq:untwisted-sector-contribution}
\end{equation}
We now want to show that in the large $N$ limit, this contribution is proportional to
\begin{equation}
  Z_{(1)^N}(\tau) = e^{\frac{N \tilde c}{12}\beta}.
  \label{eq:universal-form-untwisted-sector-contribution}
\end{equation}
up to subleading terms in $N$.
Note that this holds for all temperatures in contrast to the total partition function \eqref{eq:S_N-orbifold-partition-function} where the contribution of the untwisted sector is dominant only for $\beta > 2\pi$.
To show \eqref{eq:universal-form-untwisted-sector-contribution}, we employ the same techniques that were used in \cite{Hartman:2014oaa} to show the universality of the total partition function.
For this we define the generating function of \eqref{eq:untwisted-sector-contribution},
\begin{equation}
  \cZ(\tau) = \sum_{N=0}^\infty p^N Z_{(1)^N}(\tau) = \prod_{h,\bar h}(1-p q^{h-\tilde c/24}\bar q^{\bar h -\tilde c/24})^{-\tilde d(h,\bar h)}.
  \label{eq:generating-function-untwisted-sector-contribution}
\end{equation}
Here $\tilde d(h,\bar h)$ is the density of states of the seed theory,
\begin{equation}
  \tilde Z(\tau) = \sum_{h,\bar h} \tilde d(h,\bar h) q^{h-\tilde c/24}\bar q^{\bar h -\tilde c/24}.
\end{equation}
The right hand side of \eqref{eq:generating-function-untwisted-sector-contribution} can easily be verified by differentiating w.r.t.~$p$.
We now employ a trick of \cite{deBoer:1998us,Keller:2011xi} by rewriting $\cZ(\tau)$ in terms of $\hat p = p q^{-\tilde c/24}\bar q^{-\tilde c/24}$ and factoring out the contribution of the vacuum $h = \bar h = 0$ which we assume to be unique,
\begin{equation}
  \cZ(\tau) = \frac{1}{1-\hat p} \prod_{h,\bar h>0}(1-p q^h\bar q^{\bar h})^{-\tilde d(h,\bar h)} = \frac{1}{1-\hat p} R(\hat p).
  \label{eq:generating-function-untwisted-sector-contribution-2}
\end{equation}
Series expanding $R(\hat p) = \sum_{k=0}^\infty a_k \hat p^k$, we see that $Z_{(1)^N}(\tau) e^{-\frac{N \tilde c}{12}\beta} = \sum_{k=0}^N a_k$ and thus
\begin{equation}
  \hat Z_{(1)^\infty}(\tau) = \lim_{N \to \infty} Z_{(1)^N}(\tau) e^{-\frac{N \tilde c}{12}\beta} = \sum_{k=0}^\infty a_k = R(1).
  \label{eq:series-untwisted-sector-contribution}
\end{equation}
To prove \eqref{eq:universal-form-untwisted-sector-contribution}, we need to show that $\log\hat Z_{(1)^\infty}$ converges.
This is readily achieved by series expanding the $\log(1-q^h \bar q^{\bar h})$ terms in $\log\hat Z_{(1)^\infty}$, giving
\begin{equation}
  \log \hat Z_{(1)^\infty}(\tau) = \sum_{k=1}^\infty \frac{1}{k} ((q\bar q)^{k \tilde c/24} \tilde Z(k\tau) - 1).
\end{equation}
Due to the vacuum contribution having been factored out in \eqref{eq:generating-function-untwisted-sector-contribution-2}, $(q\bar q)^{\tilde c/24} \tilde Z(\tau) - 1$ can be bounded by (see \cite{Hartman:2014oaa})
\begin{equation}
  (q\bar q)^{\tilde c/24} \tilde Z(\tau) - 1 \leq \tilde p(\tau) q^{h_1}\bar q^{\bar h_1} e^{\frac{\tilde c}{6} \frac{4\pi^2}{\beta}},
\end{equation}
where $\tilde p(\tau)$ is some polynomial in $\tau$ and $h_1,\bar h_1$ the conformal weight of the lowest non-vacuum primary of the seed theory.
Inserting this back in \eqref{eq:series-untwisted-sector-contribution}, we see that the sum over $k$ converges and thus $\log \hat Z_{(1)^\infty}(\tau) < \infty$.
This derivation easily generalizes to the contribution of $n_m$ short cycles of length $m$ since $Z_{(m)}(\tau)$ from \eqref{eq:single-cycle-contribution} can be bounded by $Z_{(m)}(\tau) \leq p(\tau) \tilde Z(\tau/m)$ (the sum in \eqref{eq:single-cycle-contribution} can be bounded by a factor times its largest term), which implies
\begin{equation}
  Z_{(m)^{n_m}}(\tau) = e^{\frac{\tilde c}{12}\beta n_m/m},
  \label{eq:short-cycle-contribution-partition-function}
\end{equation}
up to terms subleading in $n_m \propto \cO(N)$.

\subsubsection*{Long cycles}
To show the universality of cycles cycle contributions, we start with the maximally twisted sector.
The corresponding contribution to the partition function is given by
\begin{equation}
  Z_{(N)}(\tau) = \frac{1}{N} \sum_{j=0}^{N-1} \tilde Z\left(\frac{\tau + j}{N}\right).
  \label{eq:maximally-twisted-sector-contribution}
\end{equation}
In the large $N$ limit, this scales as
\begin{equation}
  Z_{(N)}(\tau) = \frac{1}{N} e^{\frac{\tilde c}{12} N \frac{4\pi^2}{\beta}},
  \label{eq:universal-form-maximally-twisted-sector-contribution}
\end{equation}
up to $e^{-N}$ corrections.
Note that unlike for the untwisted sector, \eqref{eq:universal-form-maximally-twisted-sector-contribution} includes a prefactor scaling polynomially in $N$.
Eq.~\eqref{eq:universal-form-maximally-twisted-sector-contribution} is a straightforward consequence of the Cardy formula.
For small $j$, the argument $(\tau + j)/N$ of the seed partition function $\tilde Z$ in \eqref{eq:maximally-twisted-sector-contribution} goes to zero as $N \to \infty$, therefore $\tilde Z\left(\frac{\tau + j}{N}\right)$ is well-approximated by
\begin{equation}
  \tilde Z\left(\frac{\tau + j}{N}\right) \propto e^{\frac{\tilde c}{12} \frac{\beta N}{j^2 + (\beta/2\pi)^2}}.
\end{equation}
For large $j$ (i.e.~$j$ scaling proportional to $N$), we apply a modular transformation to get
\begin{equation}
  \tilde Z\left(\frac{\tau + j}{N}\right) = \tilde Z\left(-\frac{1}{N\tau} + \frac{\hat j}{N}\right).
\end{equation}
If $j$ scales proportional to $N$, then $\hat j$ scales proportional to $N^0$, thus we can again apply the Cardy formula to obtain
\begin{equation}
  \tilde Z\left(\frac{\tau + j}{N}\right) \propto e^{\frac{\tilde c}{12} \frac{\beta N}{1 + (\hat j\beta/2\pi)^2}}.
\end{equation}
Now $\frac{\beta}{j^2 + (\beta/2\pi)^2} \leq \frac{4\pi^2}{\beta}$ and $\frac{\beta}{1 + (\hat j\beta/2\pi)^2} \leq \frac{4\pi^2}{\beta}$, thus the leading contribution to \eqref{eq:maximally-twisted-sector-contribution} in the large $N$ limit is given by the $j=0$ term \eqref{eq:universal-form-maximally-twisted-sector-contribution}.
More generally, for multiple long cycles we obtain for the contribution of $n_m$ cycles of length $m \propto \cO(N)$ using \eqref{eq:twisted-sector-contribution},
\begin{equation}
  Z_{(m)^{n_m}}(\tau) = \frac{1}{m^{n_m} n_m!} e^{\frac{\tilde c}{12} m n_m \frac{4\pi^2}{\beta}}
  \label{eq:long-cycle-contribution-partition-function}
\end{equation}
up to terms subleading in $N$.

\section{Probability factors for the decomposition into twisted sectors}
\label{sec:probability-factors}
In this section, we calculate the probability factors $p_{\{n_m\}}$ of the decomposition \eqref{eq:thermal-density-matrix-decomposition-twisted-sectors}.
These factors determine the relation between the total entanglement entropy $S_A$ and the entanglement entropy contribution $S_{A,\{n_m\}}$ resolved w.r.t.~a single twisted sector,
\begin{equation}
  S_A = \sum_{\{n_m\}} (p_{\{n_m\}} S_{A,\{n_m\}} - p_{\{n_m\}}\log p_{\{n_m\}}).
  \label{eq:entanglement-entropy-decomposition-twisted-sectors}
\end{equation}
Such a decomposition occurs in a number of different contexts (see for instance \cite{Balian_1989,WisemanVaccaro,KlichLevitov,Lukin256,Goldstein:2017bua}), where the first term of \eqref{eq:entanglement-entropy-decomposition-twisted-sectors} has been called configurational or accessible entropy, while the second term is known as fluctuation, measurement or number entropy defined as
\begin{equation}
  S_\text{fluct.} = - \sum_{\{n_m\}} p_{\{n_m\}} \log p_{\{n_m\}}.
\end{equation}

The $p_{\{n_m\}}$ factors are defined by
\begin{equation}
  p_{\{n_m\}} = \frac{1}{Z(\tau)} Z_{(1)^{n_1}...(N)^{n_N}}(\tau).
  \label{eq:definition-probability-factors}
\end{equation}
Using the results of app.~\ref{sec:S_N-orbifold-partition-function}, we can easily evaluate \eqref{eq:definition-probability-factors} in the large $N$ limit.
For $\beta > 2\pi$, only the untwisted sector contributes, therefore $p_{\{n_m\}}$ is one for the untwisted sector and zero otherwise.
For $\beta < 2\pi$, $p_{\{n_m\}}$ vanishes if the ${\{n_m\}}$ sector contains a large number of short cycles, i.e.~if one of the $n_m$ is proportional to $N$.
Otherwise, $p_{\{n_m\}} = (\prod_m m^{n_m} n_m!)^{-1}$.
Note that $\sum_{\{n_m\}} (\prod_m m^{n_m} n_m!)^{-1} = 1$ as is appropriate for a probability distribution.
The above formulas for $p_{\{n_m\}}$ hold up to corrections of order $e^{-N}$.

The corresponding fluctuation entropy vanishes for $\beta > 2\pi$ and scales sublinearly in $N$ (roughly proportional to $\sqrt N$) for $\beta < 2\pi$.
Note that this entropy contribution drops out in the total thermal entropy, which is given by
\begin{equation}
  \begin{aligned}
    S(\beta) &= -\Tr(\rho \log \rho)\\
    &= -\sum_{\{n_m\}} \Tr(p_{\{n_m\}} \rho_{\{n_m\}} \log(p_{\{n_m\}} \rho_{\{n_m\}}))\\
    &= \sum_{\{n_m\}} p_{\{n_m\}} S_{\{n_m\}}(\beta) + S_\text{fluct.},
  \end{aligned}
\end{equation}
where $S_{\{n_m\}}(\beta)$ is the thermal entropy of the $Z_{(1)^{n_1}...(N)^{n_N}}(\tau)$ contribution to $Z(\tau)$.
This can be seen from the definition of the thermal entropy in terms of the partition function,
\begin{equation}
  S(\beta) = \log Z(\tau) - \beta \frac{\partial \log Z(\tau)}{\partial \beta} = \left\{
    \begin{aligned}
      0 &,& \beta > 2\pi\\
      \frac{2\tilde cN}{3\beta} &,& \beta < 2\pi.
    \end{aligned}\right.
\end{equation}
For $\beta > 2\pi$, the fluctuation entropy vanishes identically.
For $\beta < 2\pi$, $Z_{(1)^{n_1}...(N)^{n_N}}(\tau) = p_{\{n_m\}} e^{\frac{\tilde c N}{12} \frac{4\pi^2}{\beta}}$ and the $p_{\{n_m\}}$ prefactor cancels with the $S_\text{fluct.}$ contribution.
Since the reduced density matrix $\rho_A$ for the ordinary entanglement entropy of a spatial subregion in the full $S_N$ theory is obtained by tracing out the complement of $A$ in each of the twisted sectors separately, the entanglement entropy contains the same fluctuation entropy contribution as the thermal entropy and therefore the same cancellation happens for the entanglement entropy.

\section{Entanglement entropy for single twisted sectors}
\label{sec:entanglement-single-twisted-sectors}
This section contains a discussion about the entanglement entropy resolved for single twisted sectors $\cS = (1)^{n_1}...(N)^{n_N}$ of the $S_N$ orbifold theory.
In particular, we discuss for which subsystems $\cA$ we obtain universal results independent of the seed theory of the $S_N$ orbifold in the limit $N \to \infty$.
The most general $\cA$ consists of the union of an arbitrary number of intervals $[z^{(j)}_{m,2i-1},z^{(j)}_{m,2i}]$ in the $j$-th cycle of length $m$ where $1 \leq j \leq n_m$.
As in sec.~\ref{sec:entanglement-S_N-orbifold} we use the convention that an interval may contain the degrees of multiple fields connected by the boundary conditions if the size of the interval $|z^{(j)}_{m,2i}-z^{(j)}_{m,2i-1}|$ is larger than one.
The contribution of a particular twisted sector to the thermal partition function is determined from \eqref{eq:twisted-sector-contribution}, which decomposes into a product of contributions
\begin{equation}
  Z_{(m)^{n_m}}(\tau) = \frac{1}{n_m} \sum_{k=1}^{n_m} Z_{(m)}(k\tau) Z_{(m)^{n_m-k}}(\tau)
  \label{eq:contribution-cycles-same-length}
\end{equation}
from cycles of the same length $m$.
This product structure extends to the replica partition function, thus it is sufficient to consider the contribution $S_{\cA,\cS,m}$ to the entanglement entropy of cycles with the same length $m$ separately.
The total entanglement entropy is then obtained by summing over all $m$,
\begin{equation}
  S_{\cA,\cS} = \sum_m S_{\cA,\cS,m}.
\end{equation}

We first consider the case of a number $n_m$ of short cycles of length $m = \cO(N^0)$.
As shown in sec.~\ref{sec:S_N-orbifold-partition-function}, the contribution $Z_{(m)^{n_m}}$ of these cycles to the thermal partition function becomes universal if $n_m = \cO(N)$ (see eq.~\eqref{eq:short-cycle-contribution-partition-function}).
If we can use the methods from sec.~\ref{sec:entanglement-S_N-orbifold} to argue for vacuum block dominance of the contribution $Z_{(m)^{n_m},\alpha}$ to the replica partition function, then the universality extends to $Z_{(m)^{n_m},\alpha}$.
For a general $\cA$, this is not possible because the methods of sec.~\ref{sec:entanglement-S_N-orbifold} rely on the entangling interval to be equal in each cycle, i.e.~the $z^{(j)}_{m,i} \equiv z_{m,i}$ are equal for all $j$.
That is, treating $Z_{(m)^{n_m},\alpha}$ as the partition function of a large central charge CFT on a branched cover of the torus and then applying the decomposition of this partition function into conformal blocks only works if the branch cuts for the fields of this auxiliary CFT all lie at the same position.
If this holds, then we obtain for instance for a single interval the following universal contribution to the entanglement entropy,
\begin{equation}
  S_{\cA,\cS,m} = \frac{\tilde c}{3} n_m \log\left(\frac{1}{\epsilon}\sin\left(\frac{\pi (z_{m,2} - z_{m,1})}{m}\right)\right).
  \label{eq:short-cycle-contribution-entanglement-entropy}
\end{equation}

For long cycles of length $m = \cO(N)$, the contribution to the entanglement entropy is determined from the replica partition function $Z_{(m)^{n_m},\alpha}$ which decomposes into a number of (replica) seed partition functions with modular parameter that goes to zero as $N \to \infty$.
For a general collection of entangling intervals, this will again not give a universal result.
However, if the entangling intervals are small in the sense that $|z^{(j)}_{m,2i} - z^{(j)}_{m,2i-1}| = \cO(N^0)$, then the size of corresponding entangling intervals in the replica seed partition function goes to zero and the entangling intervals become well separated as $N \to \infty$.
For example, in the case of a single interval $[z^{(j)}_{m,1},z^{(j)}_{m,2}]$, the leading contribution to the replica partition function in limit $\alpha \to 1$ is given by $\prod_j \tilde Z^{(j)}_{\alpha}(\tau/m)$ where $\tilde Z^{(j)}_{\alpha}(\tau/m)$ includes a single entangling interval from $z^{(j)}_{m,1}/m$ to $z^{(j)}_{m,2}/m$ the size of which goes to zero as $m \to \infty$.
Contributions from other $\tilde Z_\alpha$, for instance $\tilde Z^{(j,k)}_\alpha(2\tau/m)$ which includes two entangling intervals $[z^{(j)}_{m,1}/m,z^{(j)}_{m,2}/m]$ and $[(z^{(k)}_{m,1}+\tau)/m,(z^{(k)}_{m,2}+\tau)/m]$, are suppressed by $e^{-N}$ factors as in the thermal partition function in sec.~\ref{sec:S_N-orbifold-partition-function}.
The corresponding entanglement entropy contribution is equal to
\begin{equation}
  S_{\cA,\cS,m} = \frac{\tilde c}{3} \sum_j \log\left(\frac{\beta}{2\pi\epsilon}\sinh\left(\frac{2\pi^2(z^{(j)}_{m,2}-z^{(j)}_{m,1})}{\beta}\right)\right).
  \label{eq:long-cycle-contribution-entanglement-entropy}
\end{equation}
We note that \eqref{eq:short-cycle-contribution-entanglement-entropy} and \eqref{eq:long-cycle-contribution-entanglement-entropy} are valid for all temperatures.
The phase transition observed in sec.~\ref{sec:entanglement-S_N-orbifold} comes only into play if one considers the contribution of a large number of twisted sectors including both those with long and those with short cycles.

\section{Anomalous dimensions for primaries of the $S_N$ orbifold}
\label{sec:anomalous-dimensions}
In this section, we determine bounds on the scaling with $N$ of the anomalous dimension of a general primary $\Sigma$ in an arbitrary twisted sector of the $S_N$ orbifold theory.

Let us first consider the untwisted sector.
It is easy to see that the contribution of the untwisted sector to the thermal or replica partition function is invariant under deformations to all order in perturbation theory.
Every CFT possesses an identity operator with conformal weight $h=\bar h=0$ which must be in the untwisted sector since twisting the boundary conditions incurs an energy cost.
Since conformal blocks of the identity operator give the dominant contribution to the ordinary entanglement entropy of spatial subregions under the assumption of vacuum block dominance, this implies that the ordinary entanglement entropy is invariant under deformations.

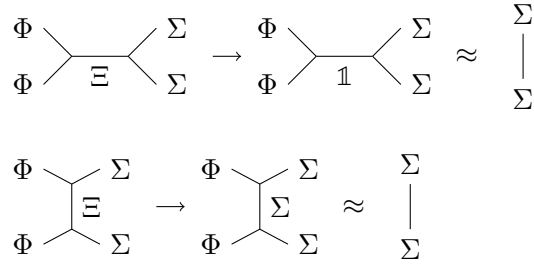
\begin{figure}
  \centering
  \begin{tikzpicture}[scale=0.5]
    \begin{scope}[shift={(0,0)}]
      \draw (-0.75,0.75) node[left] {{$\Phi$}} -- (0,0);
      \draw (-0.75,-0.75) node[left] {{$\Phi$}} -- (0,0);
      \draw (0,0) -- node[midway,below] {{$\Xi$}} (1.5,0);
      \draw (2.25,0.75) node[right] {{$\Sigma$}} -- (1.5,0);
      \draw (2.25,-0.75) node[right] {{$\Sigma$}} -- (1.5,0);
      \draw[->] (3.75,0) -- (4.5,0);
      \draw (5.75,0.75) node[left] {{$\Phi$}} -- (6.5,0);
      \draw (5.75,-0.75) node[left] {{$\Phi$}} -- (6.5,0);
      \draw (6.5,0) -- node[midway,below] {{$\mathbb{1}$}} (8,0);
      \draw (8.75,0.75) node[right] {{$\Sigma$}} -- (8,0);
      \draw (8.75,-0.75) node[right] {{$\Sigma$}} -- (8,0);
      \draw (10.5,0) node {{$\approx$}};
      \draw (12,0.6) node[above] {{$\Sigma$}} -- (12,-0.6) node[below] {{$\Sigma$}};
    \end{scope}
    \begin{scope}[shift={(0,-4)}]
      \draw (-0.75,1) node[left] {{$\Phi$}} -- (0,0.6) -- (0.75,1) node[right] {{$\Sigma$}};
      \draw (-0.75,-1) node[left] {{$\Phi$}} -- (0,-0.6) -- (0.75,-1) node[right] {{$\Sigma$}};
      \draw (0,0.6) -- node[midway,right] {{$\Xi$}} (0,-0.6);
      \draw[->] (2.25,0) -- (3,0);
      \draw (4.25,1) node[left] {{$\Phi$}} -- (5,0.6) -- (5.75,1) node[right] {{$\Sigma$}};
      \draw (4.25,-1) node[left] {{$\Phi$}} -- (5,-0.6) -- (5.75,-1) node[right] {{$\Sigma$}};
      \draw (5,0.6) -- node[midway,right] {{$\Sigma$}} (5,-0.6);
      \draw (7.5,0) node {{$\approx$}};
      \draw (9,0.6) node[above] {{$\Sigma$}} -- (9,-0.6) node[below] {{$\Sigma$}};
    \end{scope}
  \end{tikzpicture}
  \caption{The leading contribution to the anomalous dimension of $h_\Sigma$ comes from the conformal block where $\Phi\Phi$ fuse together into $\Xi = \mathbb{1}$ (upper part) or $\Phi\Sigma$ fuse together into $\Xi = \Sigma$ (lower part).
  Due to the conformal weight of $\Phi$ being of order $N^0$, these blocks are approximately equal to the two-point function $\langle \Sigma \Sigma \rangle$ in the large $N$ limit.}
  \label{fig:conformal-blocks-deformation}
\end{figure}

To determine the anomalous dimension of an operator $\Sigma$ in some other twisted sector of the $S_N$ orbifold, we insert the deformation operator $\exp\left(\lambda\int d^2w \Phi(w,\bar w)\right)$ in the two-point function $\langle \Sigma(z_1,\bar z_1)\Sigma(z_2,\bar z_2) \rangle$ and expand in the deformation parameter $\lambda$,
\begin{equation}
  \label{eq:deformed-two-point-function}
  \begin{aligned}
    \langle \Sigma(z_1,\bar z_1)\Sigma(z_2,\bar z_2) e^{\lambda\int d^2w \Phi(w,\bar w)} \rangle =~&\langle \Sigma(z_1,\bar z_1)\Sigma(z_2,\bar z_2) \rangle + \lambda \int d^2w \langle \Phi(w,\bar w) \Sigma(z_1,\bar z_1)\Sigma(z_2,\bar z_2) \rangle\\
    &+ \frac{\lambda^2}{2} \int d^2w_1 d^2w_2  \langle \Phi(w_1,\bar w_1) \Phi(w_2,\bar w_2) \Sigma(z_1,\bar z_1)\Sigma(z_2,\bar z_2) \rangle\\
    &+ \cO(\lambda^3).
  \end{aligned}
\end{equation}
Here, $\Phi(z,\bar z)$ denotes one of the exactly marginal two-cycle twist operators of the $S_N$ orbifold theory.
To determine the anomalous dimensions $h_\Sigma(\lambda)$ and $\bar h_\Sigma(\lambda)$, this is to be compared with the expansion of the two-point function in the deformed theory,
\begin{equation}
  \langle \Sigma(z_1,\bar z_1)\Sigma(z_2,\bar z_2) \rangle_\lambda = \frac{1}{(z_1-z_2)^{2h_\Sigma(\lambda)}(\bar z_1-\bar z_2)^{2\bar h_\Sigma(\lambda)}}.
\end{equation}
The $n$-th order in the series expansion of the anomalous dimension is then obtained from the $n+2$-point function of $\Sigma$ and $\Phi$ in \eqref{eq:deformed-two-point-function}.
The first order in $\lambda$ of the anomalous dimension vanishes because the OPE coefficient $C^\Phi_{\Sigma\Sigma}$ vanishes between the two-cycle twist operator $\Phi$ and any $\Sigma$.
For the second order, we expand the four-point function in \eqref{eq:deformed-two-point-function} in conformal blocks (see fig.~\ref{fig:conformal-blocks-deformation}),
\begin{equation}
  \begin{aligned}
    \langle \Phi(w_1,\bar w_1) \Phi(w_2,\bar w_2) \Sigma(z_1,\bar z_1)\Sigma(z_2,\bar z_2) \rangle = & \sum_\Xi (C^\Xi_{\Phi\Sigma})^2 \cF^\Xi_{\Phi\Sigma,\Phi\Sigma} \, \bar\cF^\Xi_{\Phi\Sigma,\Phi\Sigma}\\
    = & \sum_\Xi C^\Xi_{\Sigma\Sigma}C^\Xi_{\Phi\Phi} \cF^\Xi_{\Sigma\Sigma,\Phi\Phi} \, \bar\cF^\Xi_{\Sigma\Sigma,\Phi\Phi}.
  \end{aligned}
  \label{eq:deformation-conformal-block-decomposition}
\end{equation}
In the large $N$ limit the conformal blocks exponentiate, $\cF \sim e^{-c/6 f_\text{cl.}}$.
The fact that the semiclassical blocks $f_\text{cl.}$ increase with increasing weight $h_\Xi$ of the internal primary operator $\Xi$ \footnote{This can be seen numerically from the series expansion of the conformal blocks \cite{Hartman:2013mia}.} ensures that the leading contribution comes from the conformal block with lowest $h_\Xi$.
Note that conformal blocks in different channels exchange dominance as $z_1,z_2,w_1,w_2$ are varied, thus we need to consider all possible channels.
This is essentially a vacuum block dominance argument, only this time applied to the four-point function on the plane.
Conformal blocks for other $\Xi$ are suppressed by factors of $e^{-N}$.
The leading contribution at large $N$ is then equal to the two-point function $\langle \Sigma \Sigma \rangle$ without insertions of $\Phi(z,\bar z)$ (see fig.~\ref{fig:conformal-blocks-deformation}).

However, in general other contributions to the anomalous dimension at linear order in $N$ come from the OPE coefficients and multiplicities.
Below, we show that such contributions are absent for states in twisted sectors that consist of $n = \cO(N)$ cycles of length $\lfloor N/n \rfloor$ together with an arbitrary collection of cycles for the remaining $N - n$ fields.
Thus the conformal weight of the corresponding operators $\Sigma$ is invariant up to $\cO(N^0)$ corrections.

We now determine the combinatorical factors for the OPE coefficients contributing to the anomalous dimension of the deformed $S_N$ orbifold.
To second order in conformal perturbation theory, the relevant OPE coefficients come from the decomposition \eqref{eq:deformation-conformal-block-decomposition}.
Therefore, the task at hand is now to determine the scaling with $N$ of the OPE coefficients
\begin{equation}
  C^\Xi_{\Phi\Sigma},C^\Xi_{\Sigma\Sigma},C^\Xi_{\Phi\Phi}.
  \label{eq:OPE-coefficients}
\end{equation}
and the multiplicity factor, i.e.~the number of operators $\Xi$ with lowest weight $h_\Xi$ which are not suppressed by $e^{-N}$ factors from the conformal block.

The computation of the large $N$ scaling properties is achieved using combinatorics of the $S_N$ group.
A gauge invariant operator $\Sigma$ of the $S_N$ orbifold theory is obtained from a reference operator $\hat \Sigma$ by conjugating with elements of the $S_N$ gauge group,
\begin{equation}
  \Sigma = \frac{1}{\sqrt{A_\Sigma}}\sum_{g \in S_N} g \hat \Sigma g^{-1},
  \label{eq:S_N-gauge-invariance}
\end{equation}
where $A_\Sigma$ is a normalization factor.
The twist selection rule of the $S_N$ orbifold theory states that an n point function $\langle\hat\Sigma_1 ... \hat\Sigma_n\rangle$ with boundary conditions for the operators $\hat\Sigma_i$ determined by the $S_N$ group element $g_i$ is non-vanishing only if the product of all $g_i$ is the identity,
\begin{equation}
  \prod_i g_i = \mathbb{1}.
  \label{eq:twist-selection-rule}
\end{equation}
Therefore, the combinatorical factors we are interested in are computed by employing the twist selection rule \eqref{eq:twist-selection-rule} to count how many terms of the sum in \eqref{eq:S_N-gauge-invariance} contribute to the three-point function determining the OPE coefficients \eqref{eq:OPE-coefficients}.
See for instance \cite{Belin:2015hwa,Belin:2017nze} for related computations of OPE coefficients of the $S_N$ orbifold theory using the same techniques.

First, we determine the normalization factor $A_\Sigma$ for a general $\Sigma$ in the twisted sector determined by $\{n_m\}$ by demanding that the two-point function $\langle\Sigma(0) \Sigma(\infty)\rangle$ be equal to one.
It is easy to see that we can fix one of the two operators in $\langle\Sigma(0) \Sigma(\infty)\rangle$ to the reference operator $\hat\Sigma$, i.e.~one of the two sums over $g \in S_N$ drops out.
This yields an $N!$ factor.
The boundary conditions for the other operator must inverse to those of $\hat\Sigma$.
This happens only if the $g$ element in the remaining sum over $S_N$ is in the stabilizer subgroup $N_\Sigma$ of $\hat\Sigma$.
This subgroup is of size $|N_\Sigma| = \prod_m m^{n_m} n_m!$, yielding in total
\begin{equation}
  A_\Sigma = N! \prod_m m^{n_m} n_m!.
\end{equation}

In the next step, we determine the OPE coefficients.
We take $\Sigma$ to be in a generic $(1)^{n_1}...(N)^{n_N}$ twisted sector.
As a shorthand notation we use
\begin{equation}
  \hat C^{\Sigma_1}_{\Sigma_2\Sigma_3} \sim \sqrt{A_{\Sigma_1}A_{\Sigma_2}A_{\Sigma_3}} N! |N_{\Sigma_2}| |N_{\Sigma_3}| C^{\Sigma_1}_{\Sigma_2\Sigma_3}.
\end{equation}
The $\sim$ symbol denotes that we are considering only the combinatorical factors from the $S_N$ orbifold and not further dependencies on excitations in the seed theory.

We start with $C^\Xi_{\Phi\Sigma}$.
The action of the two-cycle twist operator $\Phi$ on $\Xi$ is to either splice together two cycles $(m1)(m2) \to (m1+m2)$ or to split apart a single cycle $(m1+m2) \to (m1)(m2)$.
If we splice together two cycles of the same length $m$, this gives a contribution $\hat C^\Xi_{\Phi\Sigma} = m^2 \frac{1}{2}(n_m+1)(n_m+2)$ since for $C^\Xi_{\Phi\Sigma}$ to be non-vanishing there must be $n_m+2$ cycles of length $m$ in $\Xi$ and thus $\frac{1}{2}(n_m+1)(n_m+2)$ possibilities to choose 2 cycles to splice together and $m$ possibilities to choose an element inside each cycle. Thus,
\begin{equation}
  C^\Xi_{\Phi\Sigma} \sim \sqrt{\frac{m^3 n_{2m} (n_m+1)(n_m+2)}{N(N-1)}}.
\end{equation}
Splicing two cycles of different lengths $m_1,m_2$ together yields $\hat C^\Xi_{\Phi\Sigma} = m_1 m_2(n_{m_1} + 1)(n_{m_2} + 1)$ with an analogous counting argument,
\begin{equation}
  C^\Xi_{\Phi\Sigma} \sim \sqrt{\frac{2 m_1 m_2 (m_1+m_2) n_{m_1+m_2} (n_{m_1}+1)(n_{m_2}+1)}{N(N-1)}}.
\end{equation}
Similarly, for splitting apart a cycle into two cycles of equal length $m$, we get $\hat C^\Xi_{\Phi\Sigma} = m (n_{2m}+1)$ for cycles with the same length $m$ since there are $n_{2m}+1$ cycles to choose and in each cycle we have $m$ possible splittings, yielding
\begin{equation}
  C^\Xi_{\Phi\Sigma} \sim \sqrt{\frac{m^3 n_m (n_m-1)(n_{2m}+1)}{N(N-1)}}.
\end{equation}
Splitting apart a cycle into two cycles of different lengths $m_1,m_2$ gives $\hat C^\Xi_{\Phi\Sigma} = (m_1 + m_2) (n_{m_1+m_2}+1)$,
\begin{equation}
  C^\Xi_{\Phi\Sigma} \sim \sqrt{\frac{2 m_1 m_2 (m_1+m_2) n_{m_1} n_{m_2} (n_{m_1+m_2}+1)}{N(N-1)}}.
\end{equation}

Next, we consider $C^\Xi_{\Sigma\Sigma}$ and $C^\Xi_{\Phi\Phi}$.
The latter OPE coefficient only allows
\begin{equation}
  \Xi \in \{(1)^N, (1)^{N-3}(3), (1)^{N-4}(2)^2\},
\end{equation}
thus we need to compute $C^\Xi_{\Sigma\Sigma}$ only for these $\Xi$.
The case $\Xi \in (1)^N$ is simple since for the identity operator all combinatorical factors drop out,
\begin{equation}
  C^\Xi_{\Sigma\Sigma} \sim C^\Xi_{\Phi\Phi} \sim 1.
\end{equation}
In the other two cases, $C^\Xi_{\Phi\Phi}$ is obtained as a special case of the $C^\Xi_{\Phi\Sigma}$ OPE coefficient derived above, giving
\begin{equation}
  \begin{aligned}
    C^\Xi_{\Phi\Phi} \sim \sqrt{\frac{12 (N-2)}{N(N-1)}} &,&& \Xi \in (1)^{N-3}(3)\\
    C^\Xi_{\Phi\Phi} \sim \sqrt{\frac{2 (N-2)(N-3)}{N(N-1)}} &,&& \Xi \in (1)^{N-4}(2)^2\\
  \end{aligned}
\end{equation}
It remains to compute $C^\Xi_{\Sigma\Sigma}$.
In order for $C^\Xi_{\Sigma\Sigma}$ to be non-vanishing $\Xi$ applied to $\Sigma$ must give an operator in the same conjugacy class as $\Sigma$.
For the case $\Xi \in (1)^{N-3}(3)$, we note that the non-trivial 3 cycle permutation in $\Xi$ applied onto a single cycle of length $m \geq 3$ again yields a cycle of length $m$ in $\binom{m}{3}$ cases.
Moreover, the 3 cycle can also act on two cycles $(m_1)(m_2)$ with $m_1 < m_2$ at the same time, giving again another two $(m_1)(m_2)$ cycles in $m_1 m_2$ cases.
Thus, for $\Xi \in (1)^{N-3}(3)$ we obtain
\begin{equation}
  C^\Xi_{\Sigma\Sigma} \sim \sqrt{\frac{3}{N(N-1)(N-2)}}\left[ \sum_{m=3}^N n_m \binom{m}{3} + \sum_{m_1=1}^N \sum_{m_2=m_1+1}^N n_{m_1}n_{m_2} m_1 m_2\right].
\end{equation}
Finally, for the $\Xi \in (1)^{N-4}(2)^2$ case, the two non-trivial 2 cycle permutations can act on either 1, 2 or 3 cycles simultaneously to give the same cycle structure again.
For one cycle $(m) \to (m)$, there are $\binom{m}{4}$ possibilities.
For two cycles $(m_1)(m_2) \to (m_1)(m_2)$ there are $m_1 m_2(m_2-2)$ possibilities for $m_1 < m_2$ and $\frac{1}{2}m^2(m-1)$ possibilities for $m_1=m_2=m$.
In the three cycle case $(m_1)(m_2)(m_1+m_2) \to (m_1)(m_2)(m_1+m_2)$, one of the 2 cycle permutations splices $(m_1)(m_2)$ together into a $(m_1+m_2)$ cycle while the other 2 cycle permutation splits apart $(m_1+m_2)$ into $(m_1)(m_2)$.
There are $m_1m_2(m_1+m_2)$ possibilities in this case if $m_1 < m_2$ and $m^3$ possibilities if $m_1=m_2=m$.
In total, we find
\begin{equation}
  C^\Xi_{\Sigma\Sigma} \sim \sqrt{\frac{8(N-4)!}{N!}}\biggl[
  \begin{aligned}[t]
    &\sum_{m=4}^N n_m \binom{m}{4} + \sum_{m_1=1}^N \sum_{m_2=m_1+1}^N n_{m_1}n_{m_2} m_1 m_2(m_2-1)\\
    &+\sum_{m=1}^N \frac{1}{4} n_m(n_m-1)m^2(m-1)\\
    &+ \sum_{m_1=1}^N \sum_{m_2=m_1+1}^{N/2-m_1} n_{m_1}n_{m_2}n_{m_1+m_2}m_1m_2(m_1+m_2)\\
    &+\sum_{m=1}^{N/4}\frac{1}{2}n_m(n_m-1)n_{2m}m^3
    \biggr].
  \end{aligned}
\end{equation}

Let us now consider an operator $\Sigma$ comprised of $n = \cO(N)$ cycles of length $\lfloor N/n \rfloor = \cO(N^0)$ and any number of cycles in the remaining $N-n$ elements.
For this operator, $n_m = n \delta_{m,\lfloor N/n \rfloor} + \cO(N^0)$.
Inserting this in the above expressions, it is easy to see that all OPE coefficients scale at most as $\cO(N^0)$.
We further note that $h_\Xi \geq h_\Sigma + \cO(N^0)$ for the $C^\Xi_{\Phi\Sigma}$ coefficients and $h_\Xi \geq 0$ for the $C^\Xi_{\Phi\Phi}C^\Xi_{\Sigma\Sigma}$ coefficients.
At the minimum value for $h_\Xi$, the conformal block reduces to the two-point function $\langle \Sigma \Sigma \rangle$ times the OPE coefficients and a multiplicity factor.
Moreover, the multiplicity factor for the $\Sigma$ in consideration is also of order $\cO(N^0)$: there are only an $\cO(N^0)$ number of twisted sectors in which $\Xi$ can lie.
Any $\cO(N)$ number of excitations on top of the ground states of these twisted sectors will also cause $h_\Xi$ to shift by an $\cO(N)$ term.
The contribution of these excited $\Xi$ is suppressed by powers of $e^{-N}$ from the conformal blocks and can be neglected.

More general $\Sigma$, however, may receive large corrections.
For instance all $\Sigma$ that are in a twisted sector containing at least one long cycle of length $m = \cO(N)$ have multiplicities and OPE coefficients that scale with $\cO(N)$.
Thus, we cannot use the above techniques to argue for the absence of $\cO(N)$ corrections to the anomalous dimension for operators in these twisted sectors.

Nevertheless, a partition or correlation function that for $\lambda=0$ is dominated by conformal blocks with operators whose scaling dimension is invariant under deformations will be dominated by the same conformal blocks in a perturbation expansion in $\lambda$.
Let for instance the partition function $Z_\lambda(\tau)$, given by
\begin{equation}
  Z_\lambda(\tau) = \sum_\Sigma \chi_{\Sigma,\lambda}(\tau) \bar\chi_{\Sigma,\lambda}(\bar\tau),
\end{equation}
be dominated by $\Sigma_0$ for $\lambda=0$,
\begin{equation}
  Z_0(\tau) \approx \chi_{\Sigma_0}(\tau) \bar\chi_{\Sigma_0}(\bar\tau).
\end{equation}
Then, the logarithm of $Z_\lambda(\tau)$ is expanded in $\lambda$ as follows,
\begin{equation}
  \log Z_\lambda(\tau) = \log Z_0(\tau) + \frac{\left.\partial_\lambda Z_\lambda(\tau)\right|_{\lambda=0}}{Z_0(\tau)}\lambda + \cO(\lambda^2),
\end{equation}
where (using that in the large $N$ limit $\chi_\Sigma(\tau) \approx e^{-\beta(h_\Sigma - c/24)}$)
\begin{equation}
  \begin{aligned}
    \left.\partial_\lambda Z_\lambda(\tau)\right|_{\lambda=0} \approx
    - &\beta \sum_\Sigma  e^{-\beta(h_\Sigma(0) + \bar h_\Sigma(0) - \frac{c}{12})}\\
    &\times\left. \partial_\lambda(h_\Sigma(\lambda) + \bar h_\Sigma(\lambda))\right|_{\lambda=0}.
  \end{aligned}
\end{equation}
Due to $h_\Sigma(0)$ scaling with $\cO(N)$ and $h_\Sigma(0) > h_{\Sigma_0}(0)$ together with analogous statements for $\bar h_\Sigma$, $\left.\partial_\lambda Z_\lambda(\tau)\right|_{\lambda=0}$ is dominated by the $\Sigma = \Sigma_0$ term non withstanding any polynomial corrections in $N$ to $h_\Sigma(\lambda)$ for $\Sigma \neq \Sigma_0$.
It is easy to see that similar arguments work for any correlation function which obeys the vacuum block dominance property for $S_N$ orbifold theory.

\bibliographystyle{JHEP}
\bibliography{bibliography}

\end{document}